\documentclass{aa}  

\usepackage{graphicx}
\usepackage{txfonts}
\usepackage[colorlinks=true, citecolor=blue]{hyperref}
\usepackage{multirow}
\usepackage{pbox}
\usepackage{booktabs}
\usepackage{color}

\newcommand{\es}{erg s$^{-1}$}

\newcommand{\kms}{km s$^{-1}$}

\newcommand{\msun}{$M_{\odot}$}
\newcommand{\mstar}{$M_{\star}$}
\newcommand{\mgas}{$M_{\rm gas}$}

\newcommand{\mdust}{$M_{\rm dust}$}
\newcommand{\kkmspc}{K~km~s$^{-1}$~pc$^2$}

\newcommand{\jykms}{Jy~km~s$^{-1}$}
\newcommand{\ci}{[C\,{\footnotesize I}]}

\newcommand{\co}{CO}
\newcommand{\cione}{[C\,{\footnotesize I}]$(^3P_1\,-\,^{3}P_0)$}
\newcommand{\citwo}{[C\,{\footnotesize I}]$(^3P_2\,-\, ^{3}P_1)$}
\newcommand{\ciplus}{[C\,{\footnotesize II}]}
\newcommand{\cofive}{CO\,$(5-4)$}
\newcommand{\coseven}{CO\,$(7-6)$}
\newcommand{\cofour}{CO\,$(4-3)$}

\newcommand{\cothree}{CO\,$(3-2)$}
\newcommand{\cotwo}{CO\,$(2-1)$}
\newcommand{\coone}{CO\,$(1-0)$}

\newcommand{\lprimecione}{$L'_{\mathrm{[C\,\scriptscriptstyle{I}\scriptstyle{]}}^3P_1\,-\, ^3P_0}$}

\newcommand{\lprimecitwo}{$L'_{\mathrm{[C\,\scriptscriptstyle{I}\scriptstyle{]}}^3P_2\,-\, ^3P_1}$}
\newcommand{\lprimecotwo}{$L'_{\rm CO(2-1)}$}

\newcommand{\lprimecofour}{$L'_{\rm CO(4-3)}$}
\newcommand{\lprimecofive}{$L'_{\rm CO(5-4)}$}
\newcommand{\lprimecoseven}{$L'_{\rm CO(7-6)}$}

\newcommand{\icofive}{$I_{\mathrm{54}}$}
\newcommand{\icotwo}{$I_{\mathrm{21}}$}

\newcommand{\lir}{$L_{\rm IR}$}

\newcommand{\lsun}{$L_{\odot}$}

\newcommand{\tdust}{$T_{\rm dust}$}
\newcommand{\zspec}{$z_{\rm spec}$}

\newcommand{\zopt}{$z_{\rm spec, opt}$}
\newcommand{\zsubmm}{$z_{\rm spec, submm}$}
\newcommand{\umean}{$\langle U \rangle$}
\newcommand{\sigmasfr}{$\Sigma_{\rm SFR}$}
\newcommand{\sigmagas}{$\Sigma_{\rm gas}$}
\newcommand{\distms}{$\Delta\mathrm{MS}$}
\newcommand{\tkin}{$T_{\rm kin}$}
\newcommand{\coratio}{$R_{52}$}

\begin{document}

   \title{CO emission in distant galaxies on and above the main sequence}

   \author{F. Valentino\inst{1,2},
          E. Daddi\inst{3},
          A. Puglisi \inst{3,4},
          G. E. Magdis \inst{1,2,5,6},
          D. Liu \inst{7},
          V. Kokorev \inst{1,2},
          I. Cortzen \inst{1,2},
          S. Madden \inst{3},
          M. Aravena \inst{8},          
          C. G\'{o}mez-Guijarro \inst{3},
          M.-Y. Lee \inst{9},
          E. Le Floc'h \inst{3},
          Y. Gao \inst{10,11},
          R. Gobat \inst{12},
          F. Bournaud \inst{3},
          H. Dannerbauer \inst{13,14},
          S. Jin \inst{13,14},
          M. E. Dickinson \inst{15},
          J. Kartaltepe \inst{16},
          \and
          D. Sanders \inst{17}}
        \institute{Cosmic Dawn Center (DAWN), Denmark\\
          \email{francesco.valentino@nbi.ku.dk}
          \and Niels Bohr Institute, University of
          Copenhagen, Lyngbyvej 2, DK-2100 Copenhagen \O, Denmark
          \and AIM, CEA, CNRS, Universit\'e Paris-Saclay, Universit\'e
          Paris Diderot, Sorbonne Paris Cit\'e, F-91191
          Gif-sur-Yvette, France
          \and Center for Extragalactic Astronomy, Durham University,
          South Road, Durham DH13LE, United Kingdom
          \and DTU-Space, Technical University of Denmark, Elektrovej
          327, DK-2800 Kgs. Lyngby, Denmark
          \and Institute for Astronomy, Astrophysics, Space
          Applications and Remote Sensing, National Observatory of
          Athens, GR-15236 Athens, Greece
          \and Max Planck Institute for Astronomy, K\"onigstuhl 17,
          D-69117 Heidelberg, Germany
          \and N\'{u}cleo de Astronom\'{i}a, Facultad de Ingenier\'{i}a y
          Ciencias, Universidad Diego Portales, Av. Ej\'{e}rcito 441,
          Santiago, Chile
          \and Korea Astronomy and Space Science Institute, 776
          Daedeokdae-ro, 34055 Daejeon, Republic of Korea
          \and  Purple Mountain Observatory \& Key Laboratory for Radio
          Astronomy, Chinese Academy of Sciences, 10 Yuanhua Road,
          Nanjing 210033, People’s Republic of China 
          \and Department of Astronomy, Xiamen University, Xiamen,
          Fujian 361005, People’s Republic of China
          \and Instituto de F\'{i}sica, Pontificia Universidad Cat\'{o}lica de
          Valpara\'{i}so, Casilla 4059, Valpara\'{i}so, Chile 
          \and Instituto de Astrof\'{i}sica de Canarias (IAC), E-38205
          La Laguna, Tenerife, Spain
          \and Universidad de La Laguna, Dpto. Astrof\'{i}sica,
          E-38206 La Laguna, Tenerife, Spain 
          \and NSF’s National Optical-Infrared Astronomy Research
          Laboratory, 950 North Cherry Avenue, Tucson, AZ 85719, USA 
          \and School of Physics and Astronomy, Rochester Institute of
          Technology, 84 Lomb Memorial Drive, Rochester NY 14623, USA 
          \and University of Hawaii, Institute for Astronomy, 2680
          Woodlawn Drive, Honolulu, HI 96822, USA}

  \authorrunning{Valentino, Daddi, Puglisi et al.}
  \titlerunning{CO emission in distant main-sequence galaxies}
   \date{Received --; accepted --}

 
  \abstract
  {We present the detection of multiple carbon monoxide CO line transitions with ALMA
    in a few tens of infrared-selected galaxies on and above the main sequence at
    $z=1.1-1.7$. We reliably detected the emission of \cofive,
    \cotwo, and \coseven+\citwo\ in 50, 33, and 13 galaxies,
    respectively, and we complemented this information with available
    \cofour\ and \cione\ fluxes for part of the sample, and modeling
    of the optical-to-mm spectral energy distribution. We retrieve a
    quasi-linear relation between \lir\ and \cofive\ or \coseven\ for
    main-sequence galaxies and starbursts, corroborating the
    hypothesis that these transitions can be used as star formation
    rate (SFR) tracers. We find the CO excitation to steadily increase as a function
    of the star formation
    efficiency (SFE), the mean intensity of the radiation field warming
    the dust (\umean), the surface density of SFR
    (\sigmasfr), and, less distinctly, with the distance from the main sequence
    (\distms).
    This adds to the tentative evidence for higher excitation of the 
    CO+\ci\ spectral line energy distribution (SLED) of 
    starburst galaxies relative to that for main-sequence objects, where the dust
    opacities play a minor role in shaping the high-$J$ CO
    transitions in our sample. However, the distinction between
    the average SLED of upper main-sequence and starburst galaxies is
    blurred, driven by a wide variety of intrinsic shapes.
    Large velocity gradient radiative transfer modeling demonstrates the
    existence of a highly excited component that elevates the CO SLED
    of high-redshift main-sequence and starbursting galaxies 
    above the typical values observed in the disk of the Milky Way. This excited component is dense and it
    encloses $\sim50$\% of the total molecular gas mass in
    main-sequence objects. We interpret the observed
    trends involving the CO excitation as mainly driven by
    a combination of
    large SFRs and compact sizes, as large \sigmasfr\
    are naturally
    connected with enhanced dense molecular gas fractions and higher
    dust and gas temperatures, due to increasing UV radiation fields, cosmic ray
    rates, and dust/gas coupling. 
    We release the full data compilation and the ancillary
    information to the community.}

   \maketitle
%

\section{Introduction}

Since its first detection in external galaxies a few decades ago, the
prominent role of the molecular gas in determining the evolution of
galaxies has been established by constantly growing
evidence, and interpreted by progressively more sophisticated theoretical
arguments \citep[e.g.,][for reviews]{young_1991, solomon_2005, carilli_2013, hodge_2020}.

On the one hand, the
detection of tens of different molecular
transitions in local molecular clouds and resolved nearby galaxies, spanning a wide range of
properties, allowed for a detailed description of the processes
regulating the physics of the interstellar medium (ISM). 
On the other hand, the observation of a handful of species and lines in
unresolved galaxies at various redshifts has been instrumental to identify the
main transformations that galaxy populations undergo with time.
In particular, it is now clear that the majority of
galaxies follows a series of scaling relations connecting their star formation
rates (SFRs), the available molecular and atomic gas reservoirs (\mgas,
$M_{\rm HI}$) and their densities and temperatures, the stellar and dust
masses (\mstar, \mdust), metallicities ($Z$), sizes, and several other
properties derived from the combination of these parameters. Two relations received
special attention in the past decade: the so-called ``main sequence'' (MS)
of star-forming galaxies, a quasi-linear and relatively tight
($\sigma\sim0.3$ dex) correlation between \mstar\ and SFR
\citep{brinchmann_2004, noeske_2007, daddi_2007, elbaz_2007,
  rodighiero_2011, whitaker_2012, speagle_2014, sargent_2014,
  schreiber_2015}; and the Schmidt-Kennicutt (SK) relation between the
surface densities of SFR and gas mass \citep[$\Sigma_{\rm SFR}-\Sigma_{\rm gas}$,][]{schmidt_1959,
  kennicutt_1998b}. Only a minor fraction of massive star-forming
galaxies, dubbed ``starbursts'' (SBs), deviate from the MS, displaying exceptional SFRs for
their \mstar\ \citep{rodighiero_2011}, and potentially larger
\sigmasfr\ at fixed \sigmagas\ \citep{daddi_2010, sargent_2014, casey_2014}. These objects are generally
related to recent merger events, at least in the local Universe, and they
can be easily spotted as bright beacons in the far-infrared and
(sub)millimeter regimes, owing to their strong dust emission 
exceeding $L_{\rm IR}>10^{11-12}$ \lsun\ \citep[(Ultra)-Luminous InfraRed
Galaxies, (U)LIRGs,][]{sanders_mirabel_1996}.

It has also become evident that the normalization of the MS rapidly increases with
redshift: distant galaxies form stars at higher paces than in the local Universe, at fixed stellar mass
\citep[e.g.,][]{daddi_2007, elbaz_2007, whitaker_2012, speagle_2014,
  schreiber_2015}. This trend could be explained by the
availability of copious molecular gas at high redshift \citep{daddi_2010,
  tacconi_2010, scoville_2017_gas, tacconi_2018, riechers_2019,
  decarli_2019, liu_2019_apj}, ultimately regulated
by the larger accretion rates from the cosmic web \citep{keres_2005,
  dekel_2009l}. Moreover, higher SFRs could be induced by an increased efficiency of star formation due
to the enhanced fragmentation in gas-rich, turbulent, and gravitationally unstable
high-redshift disks \citep{bournaud_2007, dekel_2009p,
  bournaud_2010, ceverino_2010, dekel_2014}, reflected on their clumpy
morphologies \citep{elmegreen_2007, forster-schreiber_2011,
  genzel_2011, guo_2012, guo_2015, zanella_2019}. 
IR-bright galaxies with prodigious SFRs well above the level of the MS are observed
also in the distant Universe, but their main physical driver is a matter of
debate. While a star formation efficiency
($\mathrm{SFE}=\mathrm{SFR}/M_{\rm gas}$) monotonically increasing with the distance
from the main sequence ($\Delta\mathrm{MS}=\mathrm{SFR/SFR_{MS}}$,
\citealt{genzel_2010, magdis_2012, genzel_2015, tacconi_2018, tacconi_2020}) could
naturally explain the existence of these outliers, recent works
suggest the concomitant increase of gas masses as the main driver
of the starbursting events \citep{scoville_2016,
  elbaz_2018}. In addition, if many bright
starbursting (sub)millimeter galaxies \citep[SMGs,][]{smail_1997} are
indeed merging systems as in the local Universe \citep[and references
therein]{gomez-guijarro_2018}, there are several well documented
cases of SMGs hosting orderly rotating disks at high
redshift \citep[e.g.,][]{hodge_2016, hodge_2019, drew_2020}, disputing the pure
merger scenario. The same
definition of ``starbursts'' as galaxies deviating from the main sequence has been
recently questioned with the advent of high spatial resolution
measurements of their dust and gas emission.
Compact galaxies with short depletion timescales
typical of SBs are now routinely found on the MS, being possibly
on their way to leave the sequence
\citep{barro_2017, popping_2017, elbaz_2018, gomez-guijarro_2019b,
  puglisi_2019, jimenez-andrade_2019}; or galaxies moving within the
MS scatter, due to mergers unable to efficiently boost the star formation \citep{fensch_2017} or
owing to gravitational instabilities and gas radial redistribution \citep{tacchella_2016}.\\

In this framework, a primary source of confusion stems from the
relatively limited amount of information 
available for sizable samples of high-redshift galaxies, homogeneously
selected on and above the main 
sequence. While a fine sampling of the far-IR spectral
energy distribution (SED) has now become more
accessible and a fundamental source to derive properties as
the dust mass, temperature, and luminosity \citep[e.g.,][to mention a
few recent high-resolution
surveys in the (sub)mm]{simpson_2014, scoville_2016, dunlop_2017,
  brisbin_2017, strandet_2017, franco_2018,
  zavala_2018, liu_2019_catalog, dudzeviciute_2020, simpson_2020, hodge_2020},
direct spectroscopic measurements of the cold gas in distant galaxies remain
remarkably time consuming. As a result, systematic investigations of
the gas properties focused on either one line transition in large samples of galaxies
\citep[e.g.,][]{lefevre_2019, freundlich_2019, tacconi_2018, tacconi_2020}, or
several lines in sparser samples, often biased towards the brightest
objects as (lensed) SMGs or quasars \citep[e.g.,][]{carilli_2013, bothwell_2013,
  spilker_2014, yang_2017, canameras_2018, dannerbauer_2019}. Moreover, the
spectroscopic study of normal MS galaxies at high redshift has been primarily devoted to the
determination of the total molecular gas masses and fractions via the
follow-up of low-$J$ carbon monoxide transitions (\coone\ to \cothree,
\citealt{dannerbauer_2009, daddi_2010, tacconi_2010, freundlich_2019, tacconi_2020}),
with a few exceptions (\ci, \citealt{valentino_2018, valentino_2020b,
  bourne_2019, brisbin_2019}; \ciplus, \citealt{capak_2015, zanella_2018, lefevre_2019}). Little
is known about the denser and warmer phases in distant normal disks,
but these components might hold the key to reach a deeper understanding of the galaxy
growth, being naturally associated with the star-forming gas.\\

A few pilot studies specifically targeting mid-$J$ CO transitions in
MS galaxies suggest the existence 
of significant pockets of such dense/warm molecular gas up
to $z\sim3$ \citep{daddi_2015, brisbin_2019, cassata_2020}, along with more
routinely detected large cold reservoirs traced by low-$J$ lines
\citep{dannerbauer_2009, aravena_2010, aravena_2014}. The
observed \co\ line luminosities
of moderately excited transitions as \cofive\
further correlate almost linearly with the total
IR luminosity \lir\ \citep{daddi_2015}, similarly to
what is observed for local IR bright objects and distant SMGs \citep{greve_2014,
  liu_2015, lu_2015, lu_2017, kamenetzky_2016}, suggesting their potential use
as SFR tracers. Moreover, these studies show
evidence of CO spectral line energy distributions (SLEDs) significantly more
excited in MS galaxies than what observed in the Milky Way, but less than local
(U)LIRGs and high-redshift SMGs \citep{dannerbauer_2009, daddi_2015,
  cassata_2020}. While not necessarily good proxies of the mode of
star formation (secular vs bursty) per se, (CO) SLEDs are
relevant if they can constrain the fraction of dense molecular gas
\citep{daddi_2015}, and they remain a precious source of information on the
processes heating and exciting the ISM. This has been extensively
proven by detailed studies of local galaxies, including 
spirals, ongoing mergers, starbursts, and active nuclear regions
\citep[among the others]{panuzzo_2010, papadopoulos_2010_sled,
  papadopoulos_2010_arp220, rangwala_2011,
  papadopoulos_2012, kamenetzky_2012, schirm_2014, lu_2014, wu_2015,
  mashian_2015, rosenberg_2015,
  kamenetzky_2016,
  kamenetzky_2017, lu_2017, lee_2019}. However, the study of warm and
dense molecular gas in distant MS galaxies
remains limited to a handful of objects to date.\\

Here we present the first results of a new multi-cycle campaign with the
Atacama Large Millimeter Array (ALMA), whose impressive
capabilities allowed for the survey of several species in
the span of a few minutes of on-source integration.  
We targeted multiple CO (\cotwo, \cofour, \cofive, \coseven) and neutral atomic
carbon (\cione, \citwo) line emissions in a sample of a
few tens of main-sequence and starburst galaxies at $z\sim1.3$.
Our main goal is to explore the excitation conditions of the molecular
gas in normal disks and bursty objects and to relate it with their star
formation modes, in the
attempt to cast new light on the formation scenarios mentioned above. 
In particular, we aim to explore that portion of the parameter space 
spanned by mid-/high-$J$ CO transitions in distant normal main-sequence
galaxies currently lacking a systematic coverage.
While admittedly not comparable with the wealth of information 
available for local objects and on sub-galactic scales, the
combination of new ALMA data and archival work is a first step towards
the multi-line and large statistical studies necessary to fully
unveil the origin of the trends for the normal MS systems discussed
above.\\

Part of the data has been already used in previous works.
In particular, \cite{puglisi_2019} focused on the far-IR sizes
of our sample and anticipated the blurred difference between upper
MS and SB galaxies mentioned above, revealing a significant population of
post-starburst galaxies on the main sequence. A more articulated
analysis of the role of compactness on galaxy evolution is in
preparation (A. Puglisi et al. in prep.). We have also discussed the
neutral carbon emission in two articles \citep[V18 and V20 in the rest of this
work]{valentino_2018, valentino_2020b}. Here we present the
details of the observational campaign, the target selection, the data reduction and analysis,
and we release all the measurements to the community. The present
release supersedes the previous ones and should be taken as reference.
We then explore and interpret the basic correlations among several
observables and the properties that they are connected with. 
We further investigate the excitation conditions of MS and SB galaxies by presenting the
observed high/low-$J$ CO line ratios as a function of the fundamental properties of
the sample; by attempting a simple modeling of the CO SLEDs; and by
comparing the latter with state-of-the-art simulations and analytical
predictions. 

In the main body of the manuscript we present the primary scientific
results and we provide the essential technical elements. We refer the
reader interested in finer details to the appendices.     
Unless stated otherwise, we assume a $\Lambda$CDM cosmology with
$\Omega_{\rm m} = 0.3$, $\Omega_{\rm \Lambda} = 0.7$, and $H_0 = 70$
km s$^{-1}$ Mpc$^{-1}$ and a Chabrier initial mass function
\citep[IMF,][]{chabrier_2003}. All magnitudes are expressed in the AB system.  
All the literature data have been homogenized with our conventions.
\begin{figure}
\includegraphics[width=\columnwidth]{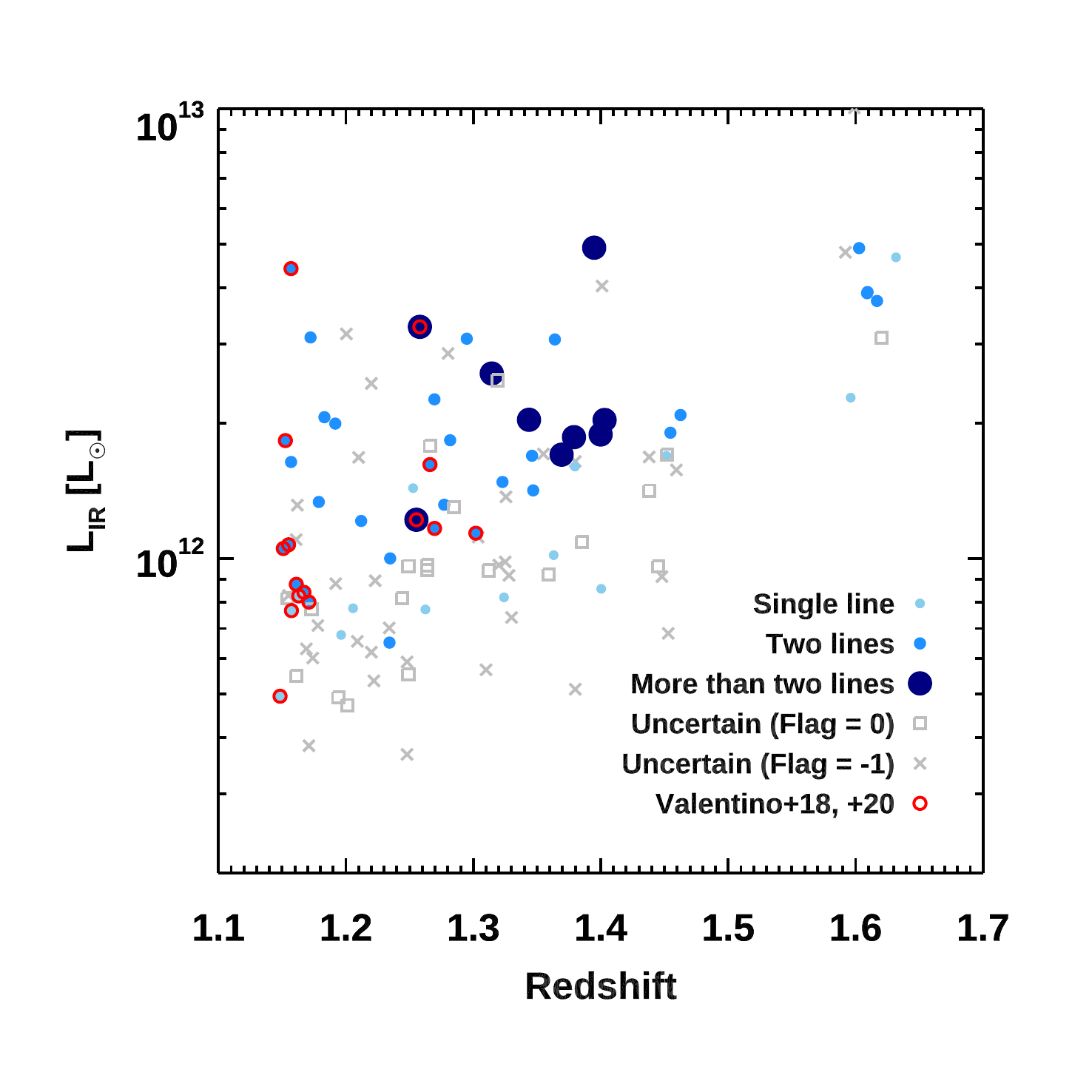}
\caption{\textbf{Survey design: the redshifts and total IR luminosities
  distributions of the primary \cofive\ sample of our
  campaign.} Gray crosses mark
  uncertain sources due to poor
  optical redshifts ($Flag=-1$). Gray open
  squares indicate uncertain upper limits on every covered line with
  trustable \zopt\ ($Flag=0$). Light blue, blue, and
  navy blue filled circles indicate reliable sources with usable
  information about one, two, or more than two transitions from this work
  ($Flag\geq0.5$). Red open
  circles mark the objects already presented in V18 and
  V20.
  Galaxies without an updated far-IR modeling based on the photometry
  in \cite{jin_2018} are not shown. The flagging classification and the definition of
  reliable and uncertain data are described in Section \ref{sec:flag_success}.}
\label{fig:design}
\end{figure}

\section{Survey description}
\label{sec:sample}
\begin{table*}
  \centering
  \caption{Summary of the multi-cycle ALMA campaign presented in this work.}
  \begin{tabular}{lcccccc}
    \toprule
    \toprule
    Epoch & Band& Transitions & Sample& $\langle t_{\rm int} \rangle$\tablefootmark{a}& Beam size& $\langle \mathrm{rms} \rangle$ \\
    & & & & \small [min]& & \small [mJy] \\
    \midrule
    Cycle 3& Band 6& \cofive & $123$& $1.6$& $0.8\arcsec$& 0.5\\
    Cycle 4& Band 3& \cotwo & $75$&   $4$& $1.4\arcsec$& 0.375\\
    Cycle 7& Band 7& \coseven, \citwo& $15$& $24$& $1.0\arcsec$& $0.25$\tablefootmark{b}\\
    \bottomrule
  \end{tabular}
  \tablefoot{
    \tablefoottext{a}{Average on-source integration time.}
    \tablefoottext{b}{Average rms.}}
  \label{tab:almaobs_recap}
\end{table*}

\subsection{The primary \cofive\ sample}
The survey was originally designed to observe \cofive\ in a
statistical sample of field IR-bright galaxies, distributed on and above the
main sequence at $z \simeq 1.1-1.7$ in the COSMOS area
\citep{scoville_2007}. The program was prepared for ALMA Cycle 3. 
The targets had available stellar mass estimates
\citep{muzzin_2013, laigle_2016}, a spectroscopic confirmation from
near-IR and optical observations from the COSMOS master catalog
(M. Salvato et al. in prep.), and a
\textit{Herschel}/PACS $100$ $\mu$m and/or 160 $\mu$m $3\sigma$
detection in the PEP catalog \citep{lutz_2011}. Initially we
considered 178 objects with a predicted \cofive\ line flux of
$I_{54}>1$ \jykms\ over 400
\kms, based on the IR luminosity and the \lir-\lprimecofive\
relation from \cite[][D15 hereafter]{daddi_2015}.
This constant flux cut corresponds to $L_{\rm IR}\gtrsim 10^{12}$
\lsun\ in the redshift interval under
consideration. We then grouped
these objects in frequency ranges within ALMA Band 6, allowing for
potential individual detections in less than two
minutes of on-source integration, while minimizing the
overheads. The final spectral sampling includes $123$ primary targets
homogeneously spread over the $z$-\lir\ space (Figure
\ref{fig:design}), with mean and median stellar masses of $M_\star =
10^{10.7}\,M_\odot$ and a 0.4 dex dispersion. 
We will refer to these galaxies as the ``primary sample'' of
our survey. Since a new
``super-deblended'' IR catalog for the COSMOS
field became available \citep{jin_2018}, a posteriori we remodeled the SED of
each object and updated the initial estimates of \lir\ (Section
\ref{sec:sed_modeling}). Twelve targets from 
the PEP catalog do not have a deblended counterpart in
\citet{jin_2018}, and, thus, do not have an updated estimate of
\lir. Moreover, two objects significantly detected in the IR do not
have a certain optical counterpart and, thus, a stellar mass estimate.
Excluding these sources, based on the new modeling 78 targets
lie on the main sequence as parameterized by \citet{sargent_2014} and 31
are classified as starbursts ($\geq3.5\times$ above the main
sequence). The threshold for the definition of starburst is arbitrary
and set in order to be consistent with V18 and V20. 
Five objects have a dominant dusty torus component in the
IR SED (see Section \ref{sec:sed_modeling}). The final distribution of
the targets in the $z$-\lir\ space with the updated IR modeling is shown in Figure
\ref{fig:design}. We note that the initial requirement of a PACS
detection steered the sample towards upper main-sequence objects and 
warm dust temperatures $T_{\rm dust}\gtrsim 30$ K.

\subsection{The \cotwo\ and \coseven\ follow-up}
 The follow-up of the \cotwo\ transition during ALMA Cycle 4 was
focused on a subsample of
75/123 objects above a constant line flux threshold of $I_{21} \geq 0.75$
\jykms. We gathered the potential targets in blocks of
frequency settings to contain the overheads, shrinking the initial
pool of galaxies with \cofive\ coverage.
Similar considerations apply for the most recent ALMA program in Cycle
7, targeting \citwo+\coseven\ in 15 galaxies with potential simultaneous visibility of
\cione. In this case, we sacrificed the flux completeness down to a constant threshold
to obtain the largest number
of multi-line measurements, adjusting the detection limit of every block
of observations to the dimmest source in each pool.\\ 

Finally, $15/123$ galaxies have at least a detection
of \cione, \cofour, \citwo, and \coseven\ from V18 and V20. In the latter, we 
operated the target selection
following a similar strategy as the one outlined here, namely by
imposing comparable \lir\ and redshift cuts. 
Figure \ref{fig:design} shows the combined information on every
targeted line available for the overall sample studied here. We will
return to the details of the detection rate below.

\section{Data and analysis}

\subsection{ALMA Observations}
\begin{figure*}
\includegraphics[width=\textwidth]{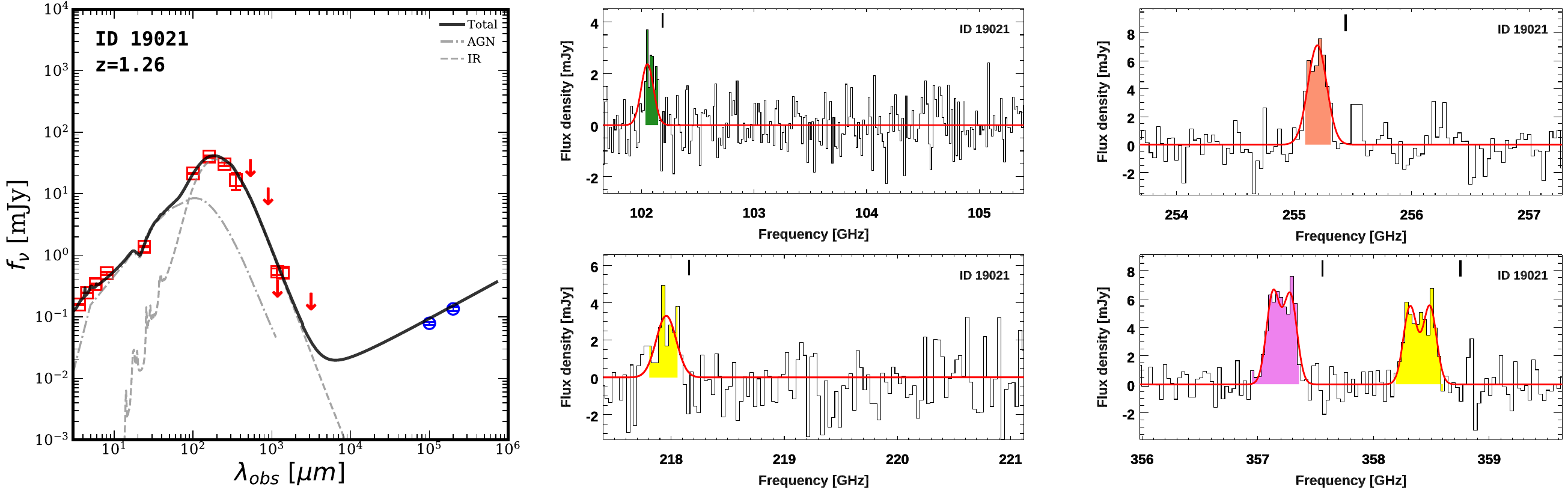}
\caption{\textbf{An example of ALMA spectra and far-IR spectral energy
    distribution.} \textit{Left:} Red open squares and arrows
  indicate detections and $3\sigma$ upper limits on the IR photometry that
  we modeled. Blue open circles indicate radio measurements, which we
  did not include in the fitting. The gray dashed and dotted-dashed
  lines mark the best-fit \cite{draine_2007} and \cite{mullaney_2011}
  templates for the star-forming and AGN components, respectively. A
  stellar template was included when modeling the IRAC
  bands, but we do not show it to avoid confusion. The
  black solid line indicates the sum of all the templates. \textit{Right:} ALMA spectra
  with detected \cotwo\ (green), \cofive\ (orange), \coseven\ (purple), and \ci\
  (yellow) transitions (\cofour\ is not available for this
  source). The solid red line indicates the best-fit
  Gaussian model. The black tick shows the expected line
  position based on the optical redshift \zopt. The full compilation
  of SEDs and spectra from which we extracted reliable line
  measurements (Section \ref{sec:flag_success}) is available in the
  online version of this article.}
\label{fig:spectra}
\end{figure*}
The primary sample of 123 targets described in Section
\ref{sec:sample} was observed in Band 6 during ALMA Cycle 3
(\#2015.1.00260.S, PI: E. Daddi). The goal of a flux density rms of 0.5 mJy, necessary
to detect a line flux of $I_{54}>1$ \jykms\ over $400$ \kms at
$>5\sigma$,  was reached in $\lesssim1.7$ minutes of integration on
source per target (Table \ref{tab:almaobs_recap}). The whole program was completed, delivering cubes
for all targets with an average beam size of $\sim0.8\arcsec$.
The subsample of 75/123 sources for the \cotwo\
follow-up was observed in Band 3 during ALMA Cycle 4
(\#2016.1.00171.S, PI: E. Daddi). With an average on-source integration of
$\sim4$ minutes, the observations matched the requested rms of $0.375$
mJy, allowing us to detect $I_{21}>0.75$ \jykms\ over $400$ \kms at
$>5\sigma$, in principle. Again, the full program was observed and
delivered, providing cubes with an average beam size of $\sim1.4\arcsec$.
Finally, 15/123 galaxies
were observed in Band 7 to detect \citwo+\coseven\ during Cycle 7
(\#2019.1.01702.S, PI: F. Valentino), reaching a flux density rms of $0.190-0.315$
mJy. We underline that the idea behind this program was to maximize the
number of sub-mm line detections per source, rather than reaching a constant flux
depth for the whole sample. Thus, the limiting rms of every block of
observations was adapted to the faintest object in each group. The
final average beam size is $\sim1.0\arcsec$.\\

\noindent
As mentioned in Section \ref{sec:sample}, 15/123 galaxies in the
primary sample have one or multiple detections of the neutral atomic
carbon \ci\ transitions, \cofour, and \coseven, as a result of an independent campaign
carried out during ALMA Cycle 4 and 6 (\#2016.1.01040.S and
\#2018.1.00635.S, PI: F. Valentino). The delivered Band 6 data have an
average beam size of $2.0\times1\arcsec.7$,
reaching a flux threshold of $\sim0.15$ \jykms per beam for a line width of $400$ \kms
(V18). The follow-up in Band 7 reached and $\mathrm{rms}$ of $0.064-0.58$ mJy,
and the resulting cubes have an average beam size of
$\sim0.9\arcsec$. We refer the reader to V18 and
V20 for more details. 

\subsection{Data reduction and spectral extraction}
\label{sec:reduction}
We reduced the data following the iterative procedure described in D15
(see also V18, V20, \citealt{coogan_2018,
  puglisi_2019, jin_2019} for reference). We calibrated the cubes using the
standard pipeline with CASA \citep{mcmullin_2007} and analyzed them
with customized scripts within GILDAS\footnote{\url{http://www.iram.fr/IRAMFR/GILDAS}} \citep{guilloteau_2000}. 
For each source, we combined all the available ALMA observations in
the \textit{uv} space allowing for an arbitrary
renormalization of the signal for all tracers. 
We then modeled each source as circular Gaussian in the \textit{uv} plane to extract the
spectrum, allowing the source position, size, and total flux per
channel to vary. Finally, the spectrum
was obtained from the fitted total fluxes per channel. This method has no obvious bias against
fitting in the image plane \citep{coogan_2018}, but it has more
flexibility in fitting parameters. 
We iteratively looked for spatially extended signal from the
$S/N$-weighted combination of all the available tracers, resorting to a point source extraction
whenever this search was not successful (D15).
In the former case, we could safely measure the size of the emitting source
and recover the total flux. When using a point source extraction, we
derived upper limits on the sizes and estimated the flux losses as
detailed in Appendix \ref{app:totalflux}. Finally, we
estimated the probability of each galaxy to be unresolved ($P_{\rm unres}$) by comparing
the $\chi^2$ of the best-fit circular Gaussian and the point source
extraction \citep{puglisi_2019}. Using a combination of
low- and high-$J$ CO transitions, \ci\
lines, and continuum emission, the sizes should be considered as
representative of the extension of the molecular gas and cold
dust in our galaxies. However, we note that the size is primarily
driven by \cofive\ and its underlying dust continuum emission
\citep[see Figure 1 of][]{puglisi_2019}. 

\subsubsection{Line and continuum flux measurements}
\label{sec:fluxmeasurements}
We blindly scanned the extracted spectra looking for potential
line emission. We did so by looking for the maximum $S/N$ computed over
progressively larger frequency windows centered on each channel. The
line flux
is then the weighted average flux density within the frequency interval
maximizing the $S/N$, times the velocity range covered by the channels
within the window, and minus the continuum emission. To estimate the
latter, we considered a $S/N$-weighted average of line-free channels
over the full spectral width ($\sim7.5$ GHz), assuming an intrinsic
power-law shape ($\propto\nu^{3.7}$). The redshift is determined from the weighted mean
of the frequencies covered by the candidate line.

To confirm the
line emission and avoid noise artifacts, we ran extensive simulations and
computed the probability for each candidate line to be spurious
($P_{\rm line}$) following the approach in \cite{jin_2019}. The
calculation provides the chance that a candidate line with a known
$S/N$, frequency, and velocity width appears in the spectrum owing to
noise fluctuations, taking into account the full velocity range
covered by the observations, the frequency sampling, and assuming a
fixed range of acceptable line FWHM. As a further check, two members of
our team visually inspected all the available spectra independently.

Once the redshift was determined, we finally remeasured the flux of each line over a fixed
velocity width, normally set by \cofive\ because of its brightness and the
widespread availability, being the primary target of our survey. Note
that we rounded the velocity width to the closest integer number of
channels allowed by the frequency resolution of each spectrum.
The adoption of identical apertures for the spectral extractions and
constant velocity ranges for the measurements allowed us to derive
consistent line ratios and depict meaningful CO SLEDs. We also note that
the redshifts and the velocity widths of the detected lines are fully
consistent, when they were left free to
vary. 

For sources without a significant flux detection, we estimated an
upper limit around the expected line position. Whenever an alternative ALMA 
line measurement was available, we adopted \zsubmm\ and the known
velocity width to set a $3\sigma$ limit of $I_{\rm line}>3 \times \mathrm{rms_{ch}}\sqrt{dv
  \Delta v}$, where $\mathrm{rms_{ch}}$ is the average noise per
channel over the line velocity width $\Delta v$, and $dv$ is the
velocity bin size. Such limits are reliable and useful, given the
exact knowledge of the redshift from the sub-mm.
When only a \zopt\ was available, we scanned the frequency
around the expected location of the line, but eventually measured only
upper limits adopting an arbitrary $\Delta v = 400$ \kms\ to be
consistent with V18 and V20. Finally, we checked our
blind flux measurements against a Gaussian parameterization of the
line emission, resulting in a $\sim10$\% flux difference due to the
different velocity widths, and fully
consistent redshift estimates, similarly to previous works
\citep[V18]{coogan_2018}. We accounted for this factor
by increasing the line fluxes from the spectral scanning by $10$\%. We note that a
blind scanning is less prone to spurious detections and it provides
more reliable estimates, whenever previous detailed knowledge of a
source is absent. Therefore, we adopted the fluxes estimated by scanning the
spectra as our final measurements.\\ 
As an example, we show the spectra of a source with multiple line detections in
Figure \ref{fig:spectra}, along with its IR SED. The whole compilation
of spectra from which we extracted reliable measurements is available
in the online version of this article.

\subsubsection{Quality flags and detection rate}
\label{sec:flag_success}
We finally classified the spectra and the \zspec\ determination for
each line by visual inspection and comparison with the optical/near-IR determination.
\begin{itemize}

  \item \textit{Flag$_{\rm line}=1$}: Secure line measurement due to low
    probability of being a false positive ($P_{\rm spurious}<0.01$)
    and/or presence of alternative lines confirming the \zsubmm,
    consistently with the optical/near-IR determination \zopt.
    \smallskip

  \item \textit{Flag$_{\rm line}=0.5$}: Secure upper limit on the line flux, given
    the presence of alternative sub-mm lines confirming \zsubmm.
    \smallskip

  \item \textit{Flag$_{\rm line}=0$}: Upper limit on the line flux, assuming a
    velocity width of $\Delta v=400$ \kms centered at the expected
    frequency based on high-quality \zopt.
    \smallskip

  \item \textit{Flag$_{\rm line}=-1$}: Undetected line and uncertain upper limit due
    to a poor quality \zopt. 
    \smallskip

  \item \textit{Flag$_{\rm line}=-99$}: Line not observed (no data).

\end{itemize}
We consider ``reliable'' the flux measurements or upper limits for
lines with \textit{Flag$_{\rm line}\geq 0.5$}, and ``uncertain'' when
\textit{Flag$_{\rm line}\leq 0$}. Adopting this scheme, we recovered
56/123 ($\sim46$\%) sources with 
 ``reliable'' \cofive\ flux estimates for the primary sample, and
 41/75 ($\sim55$\%) for \cotwo. As foreseeable for targeted
 observations, we achieved higher detection rates for \coseven\ and
 \citwo\ (13/15 detections, $\sim87$\%). Considering only sources with
 previous knowledge of \zsubmm\ rather than \zopt, the detection rate jumps to 100\%.

In Appendix \ref{app:selection} we revisit a posteriori the selection of the
targets for the \cotwo\ follow-up. This allows us to identify the
factors setting the detection rate: the quality of \zopt, lower
\lir\ than initially estimated, and intrinsically faint lines in
bright objects, in order of importance. Here we remark that the
imposition of minimum \icotwo\ and \icofive\ flux
thresholds formally biases their ratio. To account for this selection
effect, when deriving average CO SLEDs later on, we will limit the
calculation to objects that would have entered the sample of potential
\cotwo\ targets according to the revised IR modeling. 

\subsubsection{Spectroscopic redshift offset: sub-mm vs optical/near-IR}
\label{sec:zspec_comparison}
\begin{figure}
\includegraphics[width=\columnwidth]{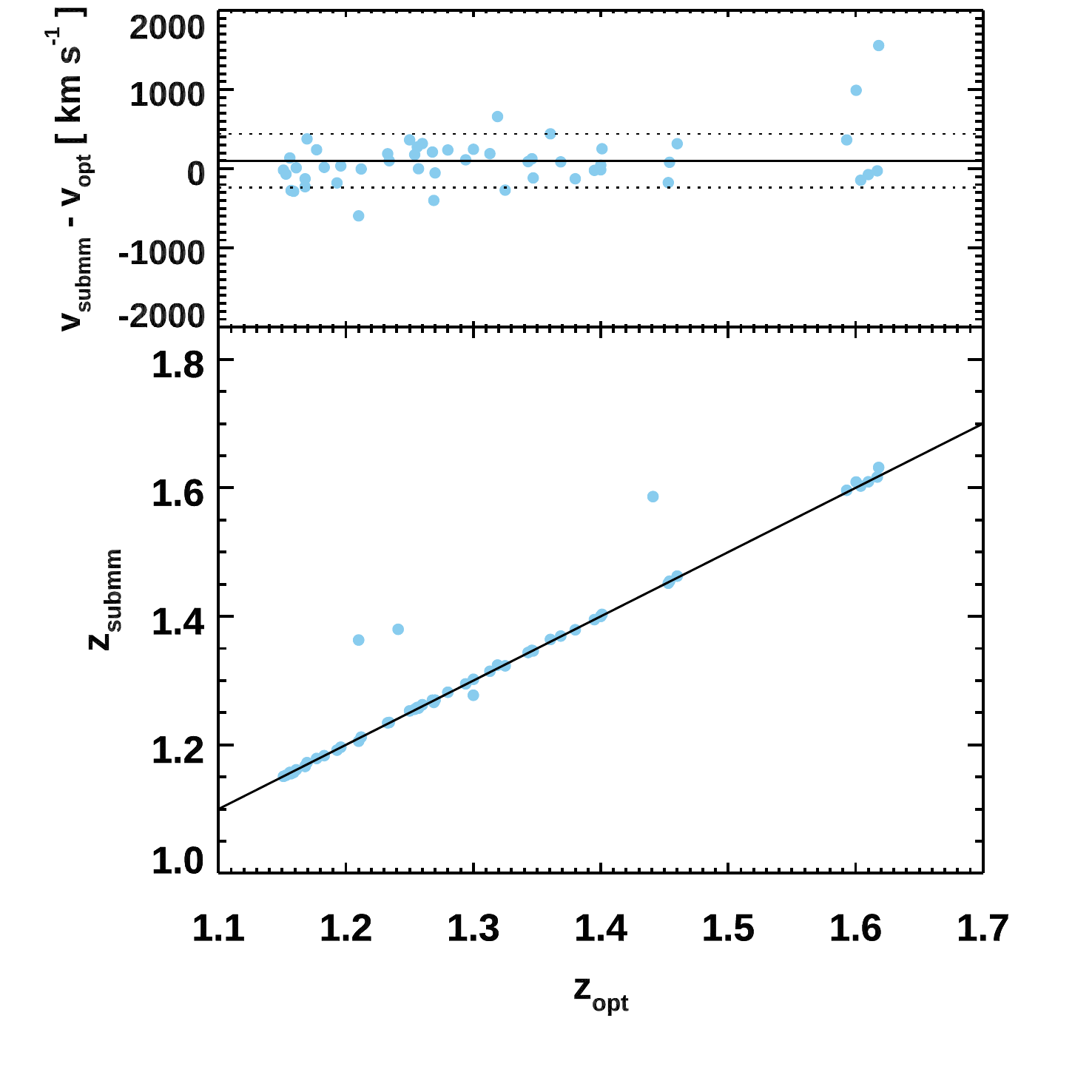}
\caption{\textbf{Spectroscopic redshift offset between submillimetric
    and optical/near-IR observations}. \textit{Bottom:} The blue
  filled circles indicate the new \zsubmm\ determinations for
  trustable emission lines detected by ALMA against the optical/near-IR \zopt\ from the
  COSMOS master compilation (M. Salvato et al. in prep.). The
  solid black line mark the locus of \zsubmm=\zopt. \textit{Top:}
  $v_{\rm submm}-v_{\rm opt}$ difference as a function of \zopt, excluding three
  catastrophic strong outliers and an ascertained
  misidentification. The solid and dotted lines mark the mean and the
  $1\sigma$ dispersion of the sample.}
\label{fig:zspec_comparison}
\end{figure}
\begin{figure}
\includegraphics[width=\columnwidth]{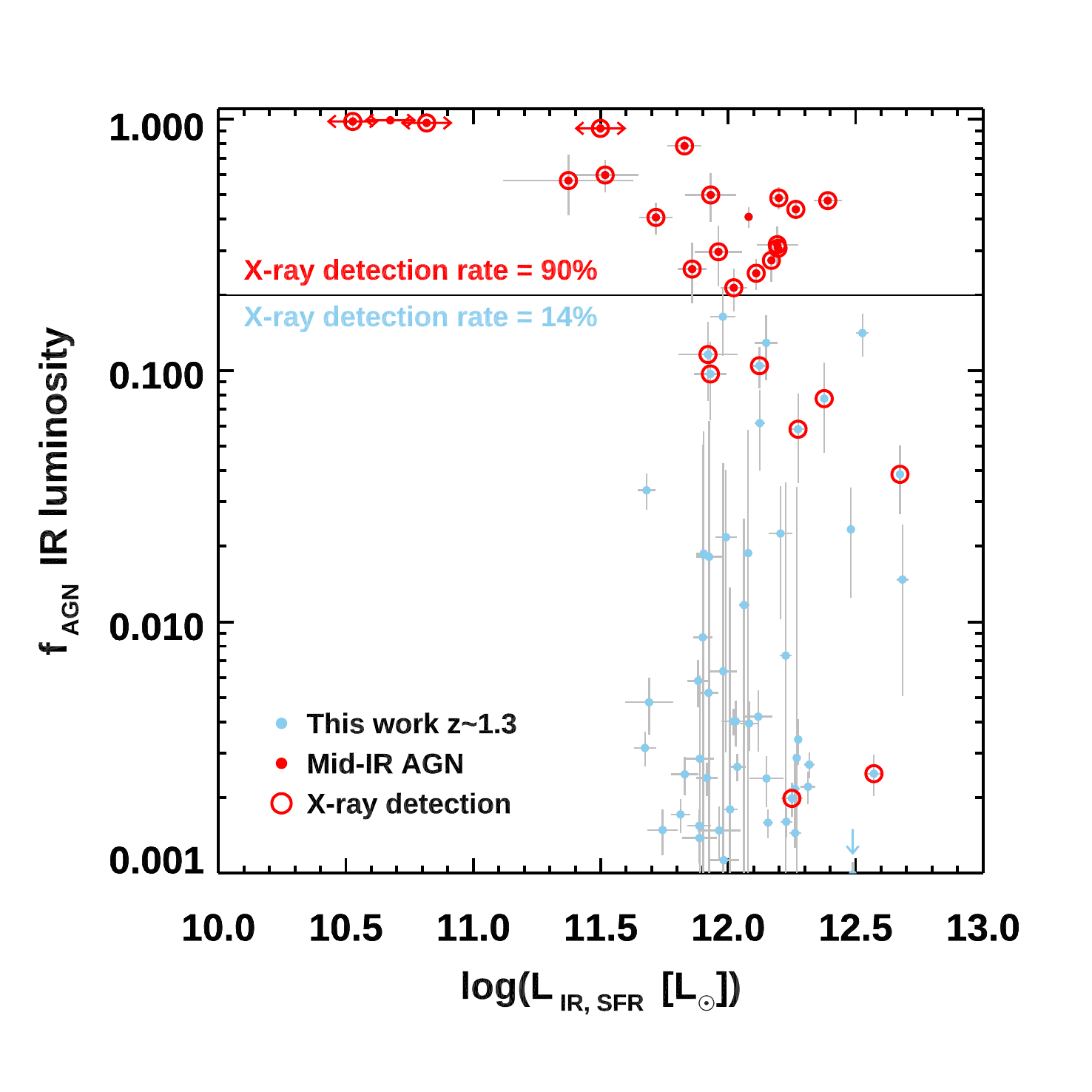}
\caption{\textbf{AGN contamination}. Fraction of the AGN contribution to
  the total IR luminosity ($f_{\rm AGN} = L_{\rm IR,
    AGN}/L_{\rm IR}$) as a function of the IR luminosity from the
  star-forming component of the SED fitting ($L_{\rm IR, SFR}$). The
  blue points show the objects with a line emission or an upper
  limit ($Flag\geq0$) and a good model of the SED. Red solid circles
  indicate mid-IR selected AGN. Red empty circles indicate
  \textit{Chandra} 2-10 keV detections with $L_{2-10\rm keV}>10^{42}$
  \es. Horizontal red arrows indicate sources with unreliable estimates of
  $L_{\rm IR, SFR}$ due to the strong AGN contamination.}
\label{fig:agn}
\end{figure}
 In Figure \ref{fig:zspec_comparison} we show the comparison between
the spectroscopic redshifts for the reliable ALMA sources and their
original \zopt\ estimate from the master compilation of the COSMOS
field (M. Salvato et al. in prep.). Excluding four sources with
ascertained significant deviations ($z_{\rm spec,submm}-z_{\rm spec,opt}>0.05$),
the ALMA and the optical/near-IR redshift 
estimates are in good agreement with mean and median $\left(z_{\rm
    spec,submm}-z_{\rm spec,opt} \right) <0.001$ and a normalized median absolute
deviation \citep{hoaglin_1983} of $\left| z_{\rm
    spec,submm}-z_{\rm spec,opt} \right| / (1+z_{\rm
  spec,submm})=0.001$. The redshift difference corresponds to a mean velocity offset
of $\langle \Delta v \mathrm{(submm-opt)}\rangle = 101\pm 46$ \kms with
a dispersion of $338$~\kms, as
measured over 54 objects. The outliers have either dubious quality or no
flags on \zopt.

\subsubsection{Serendipitous detections}
\label{ref:serendipitous}
We serendipitously found multiple sources in the dust continuum
emission maps of 12 primary targets. Four physical pairs are
spectroscopically confirmed by ALMA \citep{puglisi_2019}, while
the remaining 8/12 are detected in continuum emission only. 
Five out of 6 reliable systems in this pool are well separated and
deblended in both the $K$-band and in the far-IR. We flagged the only
other object possibly affected by blending (\#51599) and adopted the
total stellar mass and \lir\ as representative of the whole system
(see \citealt{puglisi_2019} for a possible deblending of this
source). We did not include
the confirmed deblended companions any further in the
analysis, in order to preserve the original selection. However, adding
the only object that a posteriori meets our initial criteria would not
change the main conclusions of this work.

\subsection{Infrared SED modeling}
\label{sec:sed_modeling}
We re-modeled the IR photometry of our sources from the
``super-deblended'' catalog of the COSMOS field
\citep{jin_2018} in order to derive key physical properties of the
sample, notably the total IR luminosity \lir, the dust mass \mdust,
and the mean intensity of the radiation field  
\umean. \cite{jin_2018} chose 
radio and UltraVISTA $K_s$ priors to deblend the highly confused
far-IR and sub-mm images, while performing active removal of non-relevant
priors via SED fitting with redshift information and Monte Carlo
simulations on real maps, which reduces the degeneracies and results in
well-behaved flux density uncertainties \citep{liu_2018}.
The catalog includes emission recorded by \textit{Spitzer}/MIPS at $24$
$\mu$m \citep{sanders_2007}, \textit{Herschel}/PACS \citep{lutz_2011}
and SPIRE \citep{oliver_2012} at $100-500$ $\mu$m, JCMT/SCUBA2 at 850
$\mu$m \citep{geach_2017}, ASTE/AzTEC at 1.1 mm \citep{aretxaga_2011},
and IRAM/MAMBO at 1.2 mm \citep{bertoldi_2007}, plus complementary
information at VLA/10 cm \citep[3 GHz,][]{smolcic_2017} and 21 cm
\citep[1.4 GHz,][]{schinnerer_2010}. Furthermore, we added to this list the
information on the dust continuum emission that we measured with ALMA in
the observed $0.8-3.2$ mm range.

Our modeling follows the approach of \cite{magdis_2012} and
V20. We used an expanded library of \cite{draine_2007} models and the AGN
templates from \cite{mullaney_2011} to estimate the total
IR luminosity $L_{\rm IR}(8-1000\,\mu\mathrm{m})$ -- splitting the
contribution from star-formation and dusty tori --, the dust mass
\mdust, and the intensity of the radiation field $\langle U \rangle$. The
latter is a dimensionless quantity that can be written as $\langle U
\rangle = 1/125 \times L_{\rm  IR}/M_{\rm dust}$, the constant
expressing the power absorbed per unit dust mass in a radiation field
where $\langle U \rangle = 1$ \citep{draine_2007,
  magdis_2012}. Moreover, \umean\ can be directly related to a
mass-weighted dust temperature ($\langle U
\rangle = (T_{\rm dust, mass} / 18.9\,\mathrm{K})^{6.04}$,
\citealt{magdis_2017}). The mass-weighted
$T_{\rm dust, mass}$ is $\sim10$\% lower than the light-weighted
estimate \citep{schreiber_2018_dust}. 

\subsubsection{AGN contamination}
\label{sec:agn}
 Each best-fit IR SED model bears a fraction of the total luminosity
due to dusty tori surrounding central AGN: $f_{\rm AGN} = L_{\rm IR, AGN}/L_{\rm
  IR}$, with $L_{\rm IR} = L_{\rm IR, AGN}+L_{\rm IR, SFR}$. Clearly, the ability to
detect the AGN emission critically depends on the coverage of the
mid-IR wavelengths and the intrinsic brightness of the dust
surrounding the nuclei. We, thus, consider reliably detected
the contribution from an AGN when
$f_{\mathrm{AGN}}+1\sigma_{f_\mathrm{AGN}}\geq 20$\%, while galaxies
with $f_{\rm AGN}\geq 80$\% are flagged as
AGN dominated. Estimates of $f_{\mathrm{AGN}}\lesssim 1$\% simply indicate
the absence of strong AGN components and they should not be taken at
face value. For the sake of completeness, in Figure \ref{fig:agn} we show their
statistical uncertainty $\sigma_{f_\mathrm{AGN}}$ associated with the fitting procedure.

According to our scheme, we find AGN signatures in
$\sim30$\% of the sample, as previously reported
\citep{puglisi_2019}. A similar SED-based classification largely overlaps with
\textit{Spitzer}/IRAC color criteria widely used in the literature
\citep[V20]{donley_2012}. Figure \ref{fig:agn} further
shows how our AGN scheme overlaps with the
detection rate of hard X-ray photons from
\textit{Chandra} \citep{marchesi_2016, civano_2016}. Ninety
percent of sources with $Flag\geq0$ and $f_{\rm AGN}+1\sigma_{f_\mathrm{AGN}}\geq
20$\% also have $L_{\rm 2-10\, keV}\geq 10^{42}$ \es. 
This fraction drops
to $14$\% below the fixed $f_{\rm AGN}$ threshold, which might be
considered as the rate of 
AGN contamination in our star-forming dominated sample. 
Here we limit the analysis to this classification and to the
quantification of the AGN contribution to the total IR
luminosity. We note that we will exclude the AGN-dominated galaxies from the rest
of the analysis. A specific study of the effects of the AGN on the gas excitation
is postponed to the future.
\begin{figure*}
\includegraphics[width=\textwidth]{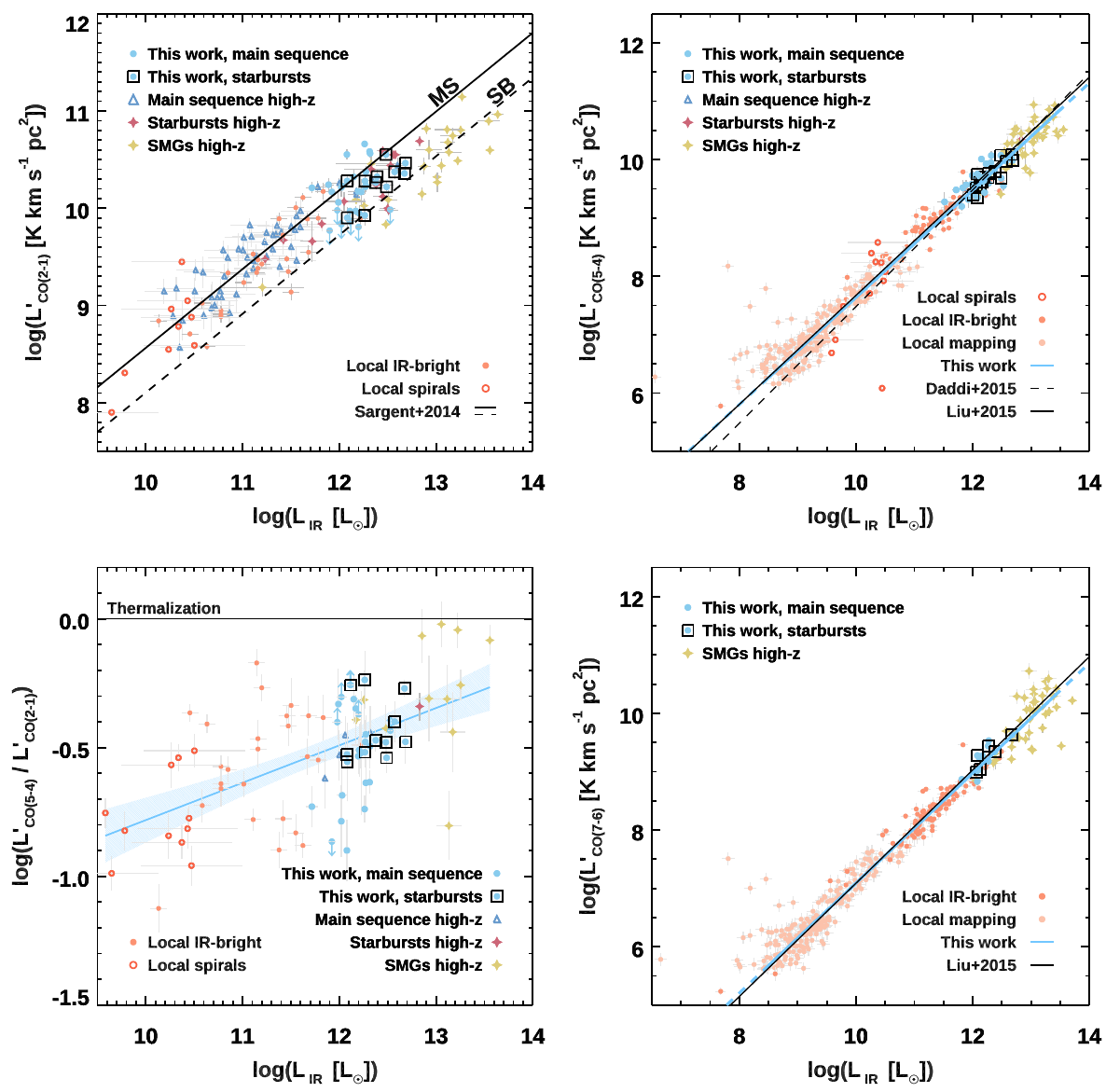}
\caption{\textbf{CO -- IR luminosities relations}. \textit{From the
    top left, clockwise:} $L'$ [\kkmspc]
  luminosities of \cotwo, \cofive, \coseven, and the \cofive/\cotwo\
  ratio as a function of the IR luminosity \lir\
  [\lsun]. Symbols are color-coded as labeled (see Section
  \ref{sec:literature} for references), where our sample of main
  sequence and starbursts ($\Delta\mathrm{MS}\geq3.5$) at $z\sim1.3$ is marked by blue filled
  circles and open black squares, respectively. The solid and dashed
  black lines show the best-fit linear models from previous works
  \citep[D15]{sargent_2014, liu_2015}. The blue lines mark our best
  model, where the dashed segments indicate its extrapolation. The
  blue shaded area in the bottom left figure shows the 95\% confidence
  interval, omitted in the remaining panels for clarity. The best-fit
  parameters and the scatter are reported in Table
  \ref{tab:scalingrelations}. We adopted an extended \lir\ range in the
  right panels to show the \lprimecofive\ and \lprimecoseven\
  luminosities obtained from mapping nearby
  objects \citep{liu_2015}, for which \cotwo\ observations on the same
  scales are not available. Note that the axes are
  inverted with respect to the canonical representation of star
  formation laws.}
\label{fig:luminosities}
\end{figure*}

\subsection{Ancillary data}
\label{sec:ancillary_data}
We took advantage of the rich ancillary data and past analysis
available for the COSMOS field by compiling stellar masses
\citep{muzzin_2013, laigle_2016} and optical/near-IR spectroscopic
redshifts (M. Salvato et al., in prep.). For sources with 
X-ray counterparts and a substantial AGN contamination, we refit the UV
to near-IR photometry using the code \textsc{CIGALE}\footnote{\url{https://cigale.lam.fr/}}
\citep{noll_2009_cigale}, self-consistently accounting for the presence of emission from
the active nuclei across the various bands as detailed in
\cite{circosta_2018}. While this provided us with a more robust estimate of
the stellar masses in presence of strong AGN, we did not find any
significant offset between \mstar\ from CIGALE and the COSMOS
catalogs for the rest of the sample.

\subsection{Literature data}
\label{sec:literature}
\begin{table}
  \centering
  \caption{CO emission from literature data.}
  \begin{tabular}{lll}
    \toprule
    \toprule
    Sample & Transition & References\\
    \midrule
    \textit{Herschel}-FTS archive & $J=1, 2$& K16\\ 
           & $J=5, 7$& L15a\\ 
    HerCULES& $J=5$& R15, L15a\\
    HERACLES& $J=2,5$& L08; L+in prep.\\
    PHIBSS-2& $J=2$& F19\\
    $BzK$ $z\sim1.5$& $J=1$& A10; A14\\
           & $J=2$& D10a\\
           & $J=3,5$& Da09, D15\\
    Starburst $z\sim1.5$& $J=2,3$& S15, S18b;\\
           & $J=5$& S18a\\ 
    SMGs& various& L15a; V18; V20\tablefootmark{a}\\
    \bottomrule
  \end{tabular}
  \tablefoot{References: K16: \citealt{kamenetzky_2016}; L15a:
    \citealt{liu_2015}; R15: \citealt{rosenberg_2015}; L08:
    \citealt{leroy_2008}; L+ in prep.:
    D. Liu et al. in prep.; F19: \citealt{freundlich_2019}; A10:
    \citealt{aravena_2010}; A14: \citealt{aravena_2014}; D10:
    \citealt{daddi_2010}; Da09: \citealt{dannerbauer_2009}; D15:
    \citealt{daddi_2015}; S15: \citealt{silverman_2015gas}; S18a,b:
    \citealt{silverman_2018sb, silverman_2018co21}; V18:
    \citealt{valentino_2018}; V20: \citealt{valentino_2020b}.\\
    \tablefoottext{a}{We refer the reader to Section
      \ref{sec:literature} of this work, V18, and V20 for detailed
      references for the SMG population.}}
  \label{tab:literature}
\end{table}
To put in context the new ALMA data, we compiled
samples from the literature (Table \ref{tab:literature}). For what concerns objects in the local
Universe, we included the local IR-bright galaxies from the full
\textit{Herschel}-FTS archive and their ancillary observations
\citep[see also V20]{liu_2015, kamenetzky_2016,lu_2017}, covering both
low- and high-$J$ CO transitions. We further added the (U)LIRGs from the HerCULES survey 
\citep{rosenberg_2015}, and the local spirals from the
HERACLES \citep{bigiel_2008, leroy_2008,
  leroy_2009} and KINGFISH surveys \citep{kennicutt_2011,
  dale_2012, dale_2017}. We note that other collections of
nearby objects with coverage of low-$J$ CO emission are available
\citep[e.g.,][]{cicone_2017, saintonge_2017, pan_2018, gao_2019}, but
we privileged galaxies with observables and properties more directly
comparable with our ALMA sample. 
\begin{table*}
  \centering
  \caption{Scaling relations involving the CO emission lines.}
  \begin{tabular}{lcccc}
    \toprule
    \toprule
    Relation$\dagger$& Slope& Intercept& Intrinsic scatter& Correlation \\
    $x, y$ & $\beta$& $\alpha$& $\sigma_{\rm int}$& $\rho$\\
    \midrule\\
    \multicolumn{5}{c}{Luminosities}\\
    \midrule
    \lir, \lprimecofive&    $0.92\pm0.01$&    $-1.55\pm0.08$&  $0.17\pm0.01$& $0.99$\\  
    \lir, \lprimecoseven&   $0.94\pm0.01$&    $-2.34\pm0.09$&  $0.16\pm0.01$& $0.99$\\  
    \lir, \coratio&    $0.15\pm0.02$&    $-2.23\pm0.28$&  $0.17\pm0.02$& $0.62$\\  
    \midrule\\
    \multicolumn{5}{c}{Distance from the main sequence}\\
    \midrule
    \distms, \lprimecotwo/\lir$^\ddagger$& $-0.34\pm0.13$& $-1.84\pm0.08$& $0.24\pm0.04$& $-0.50$\\
    \distms, \lprimecofive/\lir& $-0.01\pm0.11$& $-2.51\pm0.07$& $0.33\pm0.03$& $-0.01$\\
    \distms, \lprimecoseven/\lir$^\ddagger$& $0.08\pm0.18$& $-3.02\pm0.11$& $0.18\pm0.06$& $0.18$\\
    \distms, \coratio$^\ddagger$& $0.23\pm0.10$& $-0.64\pm0.06$& $0.13\pm0.03$& $0.55$\\
    \midrule\\
    \multicolumn{5}{c}{Drivers of the CO excitation}\\
    \midrule
    \umean, \coratio&    $0.36\pm0.05$&    $-0.95\pm0.07$&  $0.15\pm0.02$& $0.75$\\  
    SFE($Z=Z_{\rm FMR}$), \coratio&    $0.37\pm0.05$&    $-0.66\pm0.03$&  $0.11\pm0.03$& $0.88$\\  
    SFE($Z=Z_\odot$), \coratio&    $0.31\pm0.04$&    $-0.68\pm0.03$&  $0.11\pm0.02$& $0.89$\\  
    SFE($Z=Z_{\rm FMR}\lor Z_{\rm super}$), \coratio&    $0.32\pm0.04$&    $-0.69\pm0.03$&  $0.12\pm0.02$& $0.86$\\  
    \sigmasfr, \coratio&    $0.16\pm0.02$&    $-0.60\pm0.03$&  $0.14\pm0.02$& $0.82$\\  
    \bottomrule
  \end{tabular}
  \tablefoot{$\dagger$: The linear regression is applied to
    log-quantities: $\mathrm{log}(y) =
    \alpha+\beta\times\mathrm{log}(x)$.
    The model accounts for errors on both variables and censored data, adopting the Bayesian
    approach described in \cite{kelly_2007}. Double-censored data are
    not included.\\
    $\ddagger$: Fit over our sample at $z\sim1.3$ only.}
  \label{tab:scalingrelations}
\end{table*}

At higher redshifts, we incorporated the MS and SB observations from
the PHIBSS-2 survey at $z=0.5-0.8$ \citep{freundlich_2019}, the four $BzK$ galaxies in
D15 \citep{dannerbauer_2009,daddi_2010, aravena_2010, aravena_2014}, and a pool of
starbursts at $z=1.5$ \citep{silverman_2015gas, silverman_2018sb, silverman_2018co21}.
Finally, we included samples of the high-redshift sub-mm galaxies and quasars
\citep[see V18 and V20]{walter_2011,
  alaghband-zadeh_2013, aravena_2016, bothwell_2017,
  yang_2017, andreani_2018, canameras_2018, liu_2015, carilli_2013}.

For the whole compilation, we homogenized the measurements to our
assumptions (stellar IMF, cosmology, far-IR modeling). Whenever
publicly available, we refitted the far-IR photometry adopting the
same recipes as described in Section \ref{sec:sed_modeling} to avoid
systematics. This is the case for the subsample of the PHIBSS-2 survey
in COSMOS and the compilation described in V18 and V20. We used the total \lir\ 
reported in \citet{whitaker_2014} for the remaining fields covered by PHIBSS-2.
Since a similar approach to ours has been used to model the SED of $BzK$s
and SBs at $z\sim1.5$, we adopted the best-fit values reported in the original papers. 
As detailed in V20, for the local sample of IR-bright galaxies we
converted the \textit{IRAS}-based $40-400$ $\mu$m
far-IR luminosities \citep{sanders_2003} to total estimates
integrated between $8-1000$ $\mu$m as $L_{IR} = 1.2\times
L_{\rm FIR,\,40-400\,\mu m}$. We calibrated this factor on a
subsample of galaxies from the Great Observatories All-Sky LIRGs
Survey \citep[GOALS,][]{armus_2009}. 
Finally, whenever necessary, we increased by a factor of $1.5\times$ the
total \lir\ from the modified black body modeling to match the values from
\cite{draine_2007} templates \citep{magdis_2012}. 

\section{Results}
\label{sec:results}
\begin{figure*}
  \centering
  \includegraphics[width=0.9\textwidth]{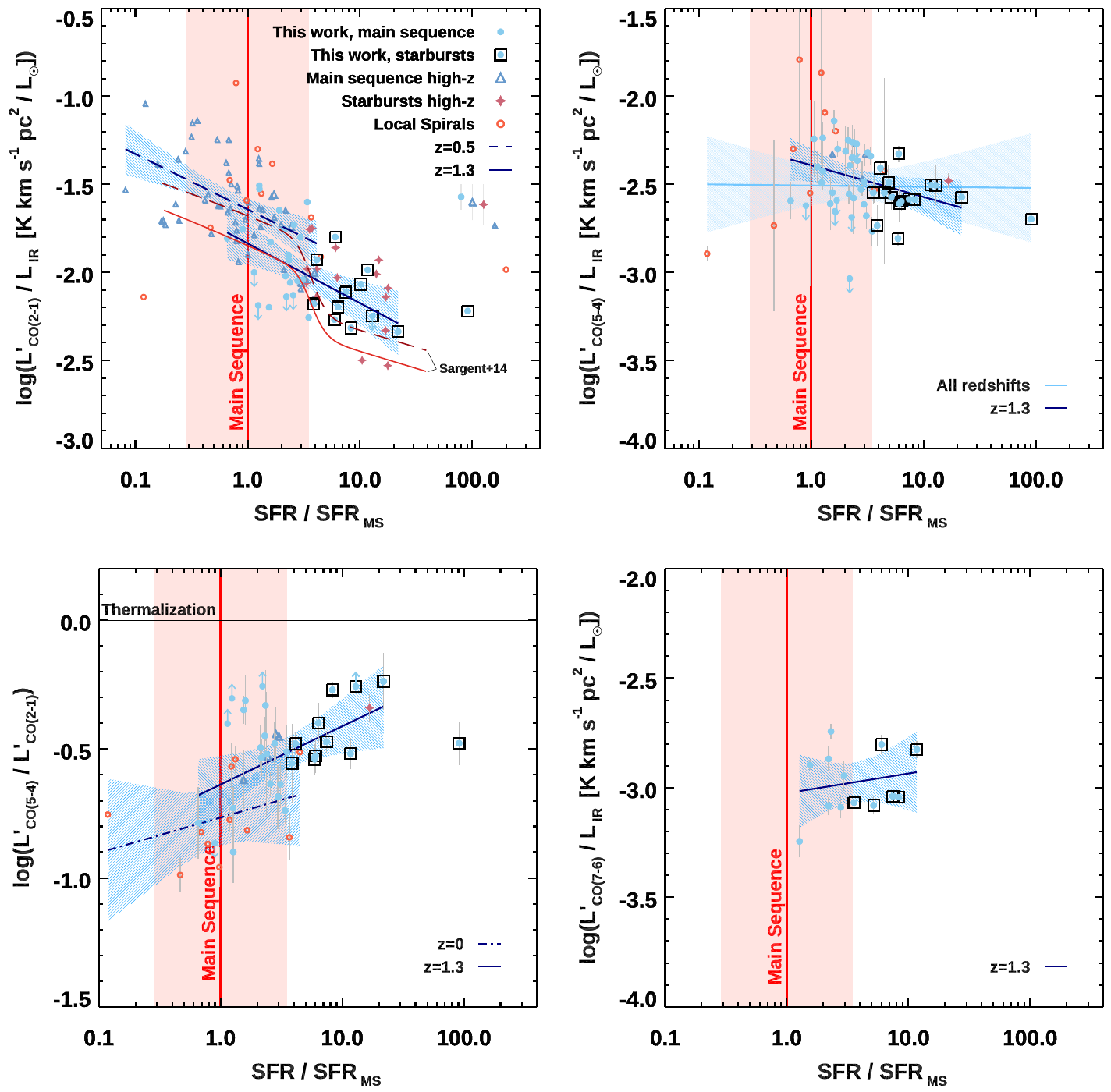}
  \caption{\textbf{CO emission on and above the main sequence}. $L'$
    [\kkmspc] / \lir\ [\lsun] for \cotwo\ (\textit{top left}), \cofive\
    (\textit{top right}), \coseven\ (\textit{bottom right}), and the 
    \lprimecofive/\lprimecotwo\ ratio (\textit{bottom left})
    as a function of the distance from the main
    sequence. In every panel, the sample of main sequence galaxies and starbursts at
    $z\sim1.3$ presented here is marked by blue filled circles and open
    black squares. Open blue triangles and purple stars indicate
    other distant MS galaxies and SBs (at $z=0.5-0.8$ from the PHIBSS-2
    survey of \cotwo,
    \citealt{freundlich_2019}; at $z\sim1.5$ from D15, \citealt{silverman_2018sb, silverman_2018co21}). 
    Open red circles signpost local spirals
    \citep[D. Liu et al. in
    prep.]{leroy_2008,kennicutt_2011,dale_2012,dale_2017}. The red
    solid line and the shaded area indicate the main sequence within
    $\|\Delta\mathrm{MS}\|<3.5$ as parameterized by
    \cite{sargent_2014}. The tracks and the shaded areas mark the best-fit
    models from the linear regression analysis and their 95\%
    confidence interval when modeling the following samples: solid dark blue
    line: our galaxies at $z\sim1.3$, excluding the strongest outlier above the main
    sequence; dashed dark blue line: PHIBSS-2
    galaxies at $z=0.5-0.8$; dotted-dashed dark blue line: local
    spirals; solid light blue line: all samples available (see
    Table \ref{tab:scalingrelations} for the parameters). The dashed and
    solid red lines in the top left panel show the tracks
    for the two-mode star formation model by \citet{sargent_2014} at
    $z=0.5$ and $1.5$ and for $M_\star=10^{10.7}\,M_\odot$, the median mass
    of both the PHIBSS-2 and our sample. The median error bars for each
    sample are displayed in the panel.}
  \label{fig:distance}
\end{figure*}

\subsection{How does the \co\ emission correlate with the infrared luminosity?}
\label{sec:lirlcorelation}
In Figure \ref{fig:luminosities} we show how low-, mid-, and 
high-$J$ \co\ $L'$ luminosities compare with the infrared
luminosities \lir, including both our new ALMA data on distant main-sequence
and starburst galaxies and the literature sample. Note that we inverted the axes
with respect to the canonical representation of the star formation laws
in order to facilitate the comparison between the various tracers and
across different figures. Here we adopted the
\lir\ corrected for the contribution of the dusty tori surrounding the
AGN, excluding those sources dominated by such contribution
($f_{\rm AGN}\geq 0.8$, Section \ref{sec:agn}) or well-known QSOs at high redshift. These
objects tend to increase the scatter of the relation, being
overluminous in the mid-IR portion of their SED.

The $L'$ luminosity of each \co\ transition strongly correlates with
\lir. The upper left panel of Figure
\ref{fig:luminosities} shows that the \cotwo\ emission in our
main-sequence and starburst galaxies is consistent with the two-mode star formation
model described in \cite{sargent_2014}, where both samples follow a
sub-linear relation with different normalizations
($\mathrm{log} (L'_{\mathrm{CO(1-0)}})=0.81\times\mathrm{log}
( L_{\rm IR} )$, corresponding to a super-linear slope of $1.23$ in the canonical
SK plane). Here we applied to the model tracks a
$L'_{\mathrm{CO(1-0)}} /L'_{\mathrm{CO(2-1)}}=1.2$ excitation correction for $BzK$-selected galaxies
\citep[D15]{daddi_2010}, but similar considerations hold
for the populations of starbursting objects and SMGs.

On the other
hand, applying a Bayesian regression analysis
(\textsc{Linmix \textunderscore err.pro}, \citealt{kelly_2007}):
\begin{equation}
\mathrm{log} \left( \frac{L'_{\mathrm{CO,} \, J}}{\mathrm{K\,
    km\, s^{-1}\, pc^2}} \right) = \alpha+\beta\times\mathrm{log} \left( \frac{L_{\rm
    IR}}{L_\odot}\right)
\end{equation}
returns $\beta\sim1$ for both $J=5, 7$ (Table
\ref{tab:scalingrelations}). Similarly to what is known for local
IR-bright and high-redshift SMGs \citep[D15]{greve_2014, liu_2015,
  lu_2015, kamenetzky_2016}, this proves that mid-/high-$J$ CO luminosities 
of distant main-sequence and starburst galaxies follow the same tight linear correlation
with the total \lir, suggesting that these transitions might be used
as SFR -- rather than molecular \mgas -- tracers. This also
suggests caution when deriving the total molecular \mgas\ from high-$J$ CO
transitions without prior knowledge of the excitation conditions. 

The modeling includes our and literature sources with
$L_{\rm IR}\geq10^{8.5}$ \lsun, reliable upper limits on the line
luminosities ($Flag=0.5$),
uncertainties on both variables, and it excludes AGN-dominated
galaxies or QSOs. However, given the large dynamical range spanned by the
observations and the small sample of bright QSOs and upper
limits, the best-fit models
are largely unaffected by their inclusion. The observed
points are similarly scattered around the best fit relations for
\cofive\ and \coseven, with an intrinsic scatter of $0.16$ dex. We
note that excluding the SMGs from the fitting reduces the scatter of
the \lir-\lprimecoseven\ relation to $0.14$ dex, but not
of the \lir-\lprimecofive\ correlation,
leaving unaltered their slopes. This is consistent with previous
determinations of the scatter of the $L'_{\rm CO, J}-L_{\rm IR}$
relations based on local IR-bright objects only, which found the $J=7$
luminosity to form the tightest relation with \lir\ \citep[e.g.,][]{liu_2015}.
On the other hand, fitting only the
high-redshift samples changes the slopes of the two relations by
$<15$\% at $<3\sigma$ significance. We note that the inclusion of the
low-redshift samples drives the difference between our regression
analysis of \lir-\lprimecofive\ and
that of D15.

\subsection{CO emission and excitation on and above the main sequence}
\label{sec:codistms}
The homogeneous IR-selection of galaxies presented above allows us to
explore the CO emission and the excitation properties over a wide
range of distances from the main sequence ($\Delta
\mathrm{MS}=\mathrm{SFR/SFR_{\rm MS}}$). This is what is shown in
Figure \ref{fig:distance}, where we report the trend of $L'/L_{\rm IR}$
ratios and a proxy for the
CO excitation ($R_{52}=$\lprimecofive/\lprimecotwo) as a function of the
position with respect to the main sequence, parameterized as in
\cite{sargent_2014}. Here we consider only the dust-obscured SFR
traced by \lir, without including the contribution from UV
emission. The latter becomes of lesser importance in massive SFGs and
at increasing redshifts, as for the samples explored here, but this
simplification does not apply to local and low \mstar\ objects. We,
thus, use galaxy samples with well
determined \lir\ for our comparison.

\subsubsection{The low-$J$ transition}
The \lprimecotwo/\lir\ ratio constantly declines
for increasing $\Delta\mathrm{MS}$, a well-known tendency generally ascribable
to the shorter depletion timescales and higher SFEs of starburst
galaxies than main-sequence objects \citep[e.g.,][]{daddi_2010b, tacconi_2010,
  genzel_2015, saintonge_2017, tacconi_2018, freundlich_2019,
  liu_2019_apj, tacconi_2020}. A linear regression analysis
confirms the existence of a meaningful anticorrelation between
\distms\ and \lprimecotwo/\lir\ (Figure \ref{fig:distance}).
However, we note that the sub-linear \lir-\lprimecotwo\ relation (Figure
\ref{fig:luminosities}), coupled with the higher \lir\ at fixed \mstar\
for distant main-sequence galaxies, introduces a redshift dependence
in the $\Delta\mathrm{MS}$-\lprimecotwo/\lir\ relation, which reflects
the increasing SFE with redshift \citep{magdis_2012}. The magnitude of this effect in the range
$z=0.5-1.5$ covered by the PHIBSS-2 and our sample can be gauged by the
shift of the tracks from the two-mode star formation model by
\cite{sargent_2014}, based on the relations shown in Figure
\ref{fig:luminosities}. The tracks are calculated for
$M_\star=10^{10.7} \,M_\odot$, the median stellar mass of both
samples, and were calibrated against the data available at that time,
which did not include a significant population of starbursts. 
When fitting separately our $z\sim1.3$ and the PHIBSS-2 samples at
$z=0.5-0.8$, we retrieve a similar displacement. The slopes are similar
and consistent with the shallow increase of the SFE along the main
sequence reported by
\cite{sargent_2014}, but we do not detect any abrupt change when
entering the starburst regime.

\subsubsection{The mid-/high-$J$ \co\ transitions}
On the other hand, both the \lprimecofive/\lir\ and \lprimecoseven/\lir\ ratios are constant as a
function of $\Delta\mathrm{MS}$ (Table
\ref{tab:scalingrelations}), following the linear \lir$-L'_{J=5,7}$ correlation shown in
Figure \ref{fig:luminosities}. This strengthens the idea that mid-
and high-$J$ transitions do trace the SFR, rather than the
total molecular \mgas\ in galaxies, and they do so independently of
their stellar mass and redshift, within the parameter space of massive
and metal-rich objects spanned by the observations presented
here. Then, the
\coratio\ ratio naturally rises as the
distance from the main sequence increases: the CO emission in
starburst galaxies appears more excited than in main-sequence objects at
similar stellar masses and redshifts (Table
\ref{tab:scalingrelations}). As for \lprimecotwo/\lir, this relation
is expected to evolve with redshift, mimicking the decrease of SFE
over cosmic time. A separate fit for the local and the $z\sim1.3$
galaxies seems to suggest this evolution, even if the small statistics of objects with
both \cotwo\ and \cofive\ lines available, especially on the lower
main sequence, makes the $\Delta\mathrm{MS}-R_{52}$ trend more uncertain.
The correlations are robust against the
exclusion of the strongest outliers (Figure
\ref{fig:distance}). We note that the presence of sources on the
MS with large ratios blurs the difference with SBs \citep[see also][]{puglisi_2019}. 
A diversity of gas excitations conditions even among MS galaxies is
evident.

\subsection{Main physical drivers of the CO excitation}
\label{sec:codrivers}
\begin{figure*}
\includegraphics[width=\textwidth]{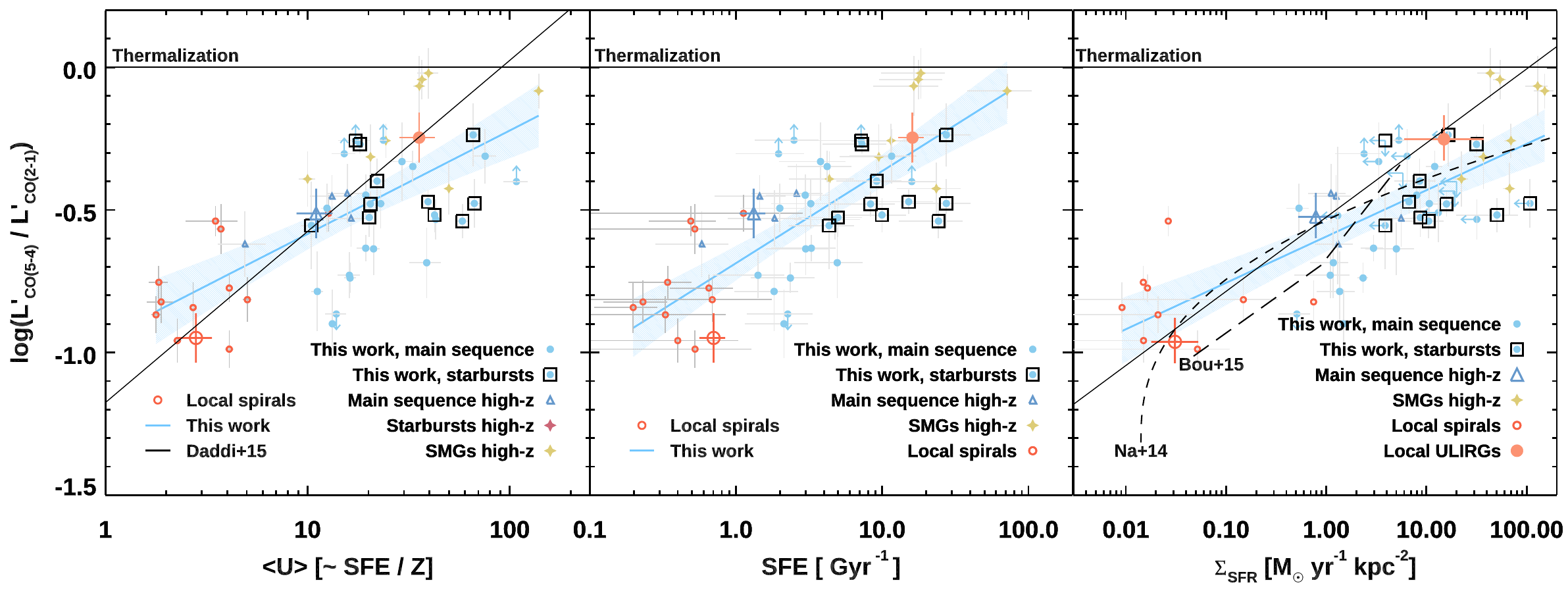}
\caption{\textbf{Physical drivers of the CO excitation.} $L'$ [\kkmspc]
  luminosity ratio between \cofive\ and \cotwo\ as a function of
  $\langle U\rangle$ from SED modeling (\textit{left}), SFE
  (\textit{center}), and $\Sigma_{\rm SFR}$
  (\textit{right}). Symbols are color-coded as labeled (see Section
  \ref{sec:literature} for references), where our sample of main
  sequence and starbursts at $z\sim1.3$ is marked by blue filled
  circles and open black squares, respectively. The solid blue line
  and the shaded area
  mark the best-fit model from the linear regression analysis and its 95\%
  confidence interval (see Table \ref{tab:scalingrelations} for the
  parameters). The solid black line shows the best-fit model from D15,
  based on the average values for local spirals, ULIRGs, and $BzK$s at
  $z=1.5$ (large open red circle, filled red circle, and open blue
  triangle, respectively). The short- and long-dashed black lines in
  the right panel indicate the predicted trends from the simulations
  by \cite{narayanan_2014} and \cite{bournaud_2015}, respectively.}
\label{fig:drivers}
\end{figure*}
We now explore the relation between a proxy of the CO
excitation conditions -- the \coratio\ ratio -- and a
few physical quantities potentially steering the molecule's
excitation: the star formation efficiency ($\mathrm{SFE=SFR}/M_{\rm
  gas}$), its combination with the metallicity as probed by the mean interstellar
  radiation field intensity heating the dust ($\langle U \rangle \propto \mathrm{SFE}/Z$), and the
star formation surface density (\sigmasfr, Figure
\ref{fig:drivers}). Since we cannot spatially resolve the CO emission in our
targets over many beams, this comparison applies to global galaxy
scales. By complementing our measurements with the existing literature,
we can span a wide interval of redshifts, masses, SFRs, and ISM
conditions. The addition of a few tens of main-sequence and
starburst galaxies further allows us to derive the average
trends among different quantities and to explore their scatter.

\subsubsection{The star formation efficiency and the mean interstellar
  radiation field intensity}
\label{sec:drivers:umean}
By homogeneously modeling the far-IR SEDs of our sample and objects from the literature
(Section \ref{sec:sed_modeling}), we retrieve sub-linear correlations
between \umean\ or SFE and \lprimecofive/\lprimecotwo, as previously reported by D15. For our own
sample of ALMA detections and reliable upper limits, we calculated SFE
by converting the \mdust\ from the SED fitting with
\cite{draine_2007} models into \mgas\ applying
a metallicity dependent gas-to-dust ratio $\mathrm{log}(\delta_{\rm
  GDR}(Z))=10.54-0.99\times(12+\mathrm{log(O/H)})$ \citep{magdis_2012},
assuming that galaxies on the main sequence follow a fundamental mass-metallicity
relation \citep[FMR,][]{mannucci_2010}. To be consistent with our previous
work, we then assumed that starburst galaxies
have supersolar metallicities ($\delta_{\rm GDR}\sim30$, while for
reference $\delta_{\rm GDR}\sim85$ for $Z=Z_\odot$,
\citealt{magdis_2012}, see also \citealt{puglisi_2017}). We factored the 0.2 dex dispersion of the assumed
mass-metallicity relation into the uncertainty of SFE. As a consistency
check, we also modeled the SFE assuming
a $\delta_{\rm GDR}(Z)$ with $Z=Z_{\rm FMR}(M_\star, \mathrm{SFR})$
and $Z=Z_\odot$ for every galaxy in our sample, on and above the main
sequence, retrieving consistent results within the uncertainties. 
We applied the same prescriptions to the literature data, considering SMGs as 
starbursting galaxies. This exacerbates the differences among
observables (or, at least, parameters closer to measurements) when
comparing starbursts and main-sequence galaxies. We warn the reader
that these are well
documented uncertainties on the
use of dust as a molecular gas tracer \citep{magdis_2012, groves_2015,
  scoville_2016}, but similar considerations
apply to \co\ and its $\alpha_{\rm CO}$ conversion factor \citep{bolatto_2013}. The choice
of using dust instead of \cotwo\ to derive \mgas\ was dictated by the attempt to reduce the correlation
with the quantity under scrutiny, \coratio.\\ 

The degeneracy on SFE driven by the $\delta_{\rm GDR}$ is partially
alleviated when using \umean\ (Figure \ref{fig:drivers}).
\umean\ carries similar information to SFE, mapped through an
assumption on the metallicity ($\langle U
\rangle \propto \mathrm{SFE}/Z$). However, it does not imply an
unknown conversion, since $\langle U \rangle \sim L_{\rm
    IR}/M_{\rm dust}$, while still prone to assumptions as the
optical depth of the dust emission (see Section \ref{sec:opacity}).
As clear from Figure \ref{fig:drivers}, starbursts and SMGs tend to
display larger \umean\ and CO line ratios than main-sequence galaxies
and local spirals, but the distinction in \umean\ is more blurred than
in SFE. For reference, we also
show the mean location of local spirals, ULIRGs, and $BzK$s at
$z\sim1.5$ from D15. The linear regression analysis in the
logarithmic space (Table
\ref{tab:scalingrelations}) returns sub-linear trends as in D15, but pointing
towards a $1.7\times$ smaller slope and with a larger intrinsic
scatter ($0.11-0.15$ dex, Table
\ref{tab:scalingrelations}).

\subsubsection{The SFR surface density}
\label{sec:drivers:sigmasfr}
The right panel of Figure \ref{fig:drivers} shows the relation between
$\Sigma_{\rm SFR}$ and \coratio. For each object, we computed
$\Sigma_{\rm SFR}=\mathrm{SFR}/(2\pi R^2)$,
where $R$ is a representative value of the galaxy radius. The latter
is rather arbitrary and it depends on the chosen tracer, the 
depth, resolution, and wavelength of the observations. Here we adopted
the ALMA sizes from circular Gaussian fitting for our sample, assuming
$R=\mathrm{FWHM}/2$. As mentioned in Section \ref{sec:reduction}, this estimate combines all the
available lines and continuum measurements, resulting in a size representative of
the dust and gas content of each galaxy \citep{puglisi_2019}. We
further recomputed the \sigmasfr\ for the $BzK$
galaxies in D15, using the Gaussian best-fit results of the rest-frame UV
observations to be consistent with our estimates. For the SPT-SMGs, we
used the sizes of \cite{spilker_2016}, while we employed the 1.4
GHz radio measurements in \cite{liu_gao_greve_2015} for the local
spirals. For reference, we also show the mean values for the $BzK$
galaxies, the local spirals, and ULIRGs as in D15.
The best-fit model to the observed points returns a $60$\% flatter
slope than in D15 (Table \ref{tab:scalingrelations}), but the trends are
qualitatively similar. We restate that the choice of the tracer, the resolution, and depth
of the observations play a major role in setting the exact values of
the slope and intercept in our simple linear model, which should be thus
taken with a grain of salt. This is particularly true for
spatially resolved local objects, where we attempted to replicate the
global, galaxy-scale measurements that can be obtained for distant
objects. The observed data points in Figure \ref{fig:drivers}
qualitatively agree with the simulations by \cite{narayanan_2014} and
\cite{bournaud_2015}, and they support the validity of \sigmasfr\ as a good
proxy for the gas conditions in galaxies. The total SFR
is a worse predictor of the gas excitation conditions \citep{lu_2014, kamenetzky_2016}, since it
does not correlate with the density and temperature probability
distribution functions in clouds
\citep{narayanan_2014}. Interestingly, this seems to be partially confirmed by
the linear regression analysis we applied here (Table
\ref{tab:scalingrelations}): when modeling \coratio\
as a function of \lir\ ($\propto\mathrm{SFR}$, Figure
\ref{fig:luminosities}) and \sigmasfr, we do find similar slopes, but
a larger linear coefficient $\rho$ for \sigmasfr\ than for \lir. However, \lir\ alone
does correlate with the CO line luminosity ratio.
\begin{figure*}
\includegraphics[width=\textwidth]{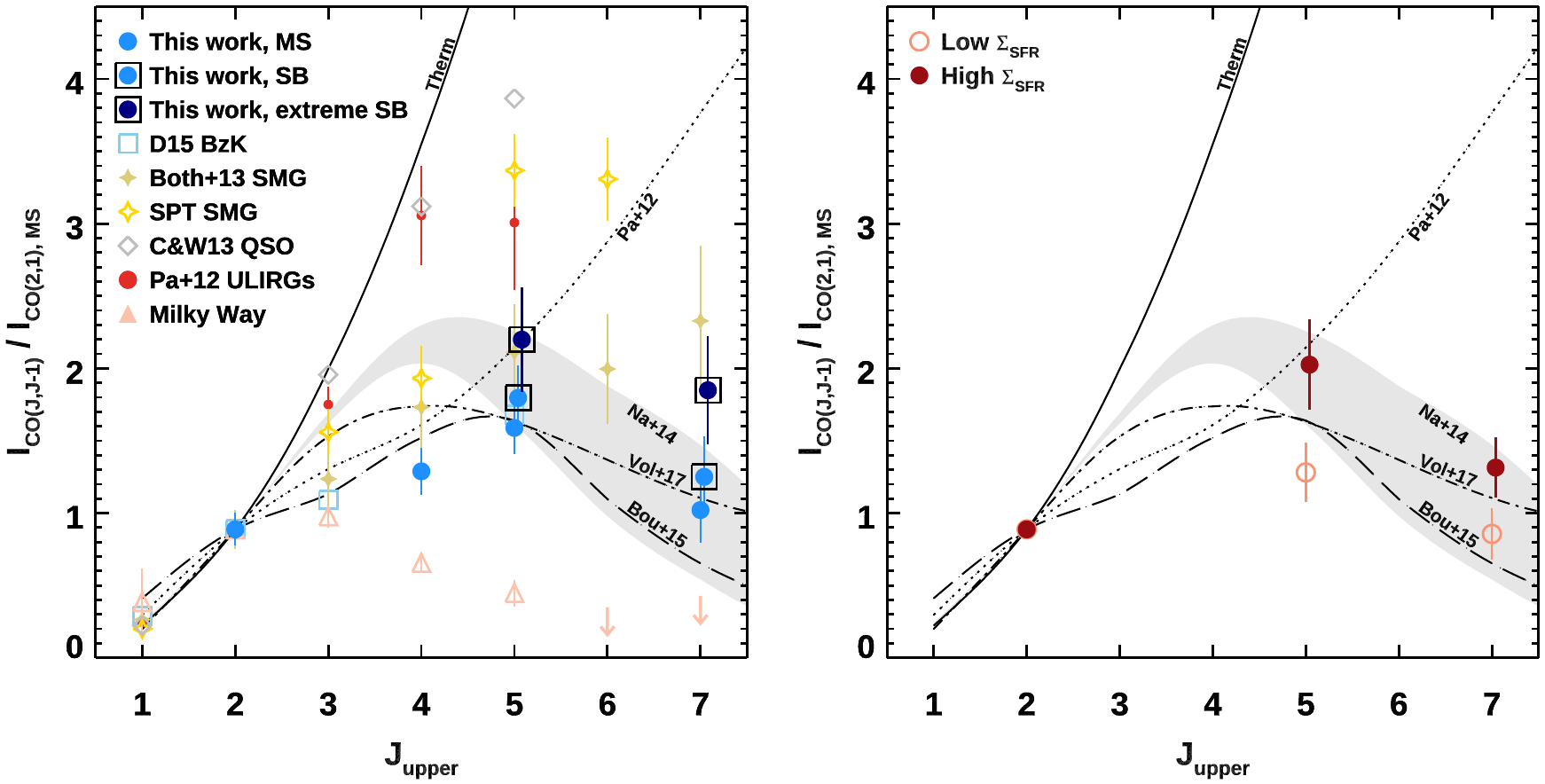}
\caption{\textbf{CO excitation ladder.} \textit{Left:}
  Average CO SLEDs of our samples compared with results from the
  literature. Each SLED is normalized to the
    mean \cotwo\ flux of main-sequence galaxies at $z\sim1.3$. Blue filled
  circles and open black squares indicate the mean SLEDs of galaxies
  on the main sequence, starbursts ($\Delta\mathrm{MS}\geq3.5$), and
  extreme starbursts ($\Delta\mathrm{MS}\geq7$), as labeled.
  The literature samples include the
  $BzK$ galaxies from D15 (open light blue squares); the average SMGs
  from \cite{bothwell_2013} (filled green stars) and the SPT-selected
  ones from \cite{spilker_2014} (open golden stars); high-redshift
  QSOs from \cite{carilli_2013} (open gray diamonds); local ULIRGs
  from \cite{papadopoulos_2012} (red filled circles); the Milky Way
  from \cite{fixsen_1999} (open pink triangles). The upper limits are
  at $3\sigma$ significance. The solid line
  shows the line ratios for a fully thermalized case. The gray shaded
  area marks the model by
  \cite{narayanan_2014} for unresolved
  observations within $\Sigma_{\rm SFR} = 1-10\,M_\odot\,\mathrm{yr^{-1}\,kpc^2}$. The 
  long dashed line indicates the simulations from
  \cite{bournaud_2015}. The dotted line tracks the empirical model by
  \cite{papadopoulos_2012}. The dashed-dotted line points at the
  analytical model by \cite{vollmer_2017}. \textit{Right}: Mean CO SLEDs
  for a subsample of objects with detections of \cotwo, \cofive,
  and estimates of \lir\ and FIR sizes, split at the median value of
  \sigmasfr. Light empty and dark filled red circles indicate the low
  and high \sigmasfr\ subsamples. Both SLEDs are normalized to the \cotwo\
  flux of the main-sequence sample as in the left panel.}
\label{fig:sled}
\end{figure*}

\subsection{The CO spectral line energy distribution of distant main-sequence galaxies}
\label{sec:cosled}
Given the large number of galaxies with available and reliable
information on \cotwo\ and \cofive, in the previous sections we used
the ratio of these two lines as a proxy for the CO
excitation. However, information of \cofour\ and \coseven\ is now
available for a subsample of distant main-sequence galaxies, which can be used
to constrain their full CO SLED. 
In Figure \ref{fig:sled} we show the average
SLEDs for main-sequence and starbursting sources, and we compare them with
a selection from the literature, representative of several different
galaxy populations. The latter range from the Milky Way \citep{fixsen_1999} and
local ULIRGs (\citealt{papadopoulos_2012}; see also \citealt{lu_2014}
and \citealt{kamenetzky_2016} for extended libraries of local
IR-bright objects), to $BzK$-selected star-forming
objects (D15), variously selected high-redshift SMGs
\citep{bothwell_2013, spilker_2014}, and QSOs \citep{carilli_2013}. The
mean and median $L'$ luminosities for our new SLEDs of main-sequence and
starburst galaxies, along with their uncertainties, are reported in
Table \ref{tab:sled}. We computed these values for the objects meeting
the requirements for a potential \cotwo\ follow-up, based on the
updated IR photometry.
These galaxies largely overlap with the sample that was effectively observed and
reliably characterized (Appendix \ref{app:selection}) and restraining
the analysis to the latter does
not affect the results of the following sections, while extending the
calculation to all the galaxies with \cofive\ would artificially decrease the observed
ratios. The imposed condition further implies similar \lir\ for the
galaxies entering the analysis. We note
that the median \lir\ for the main-sequence objects with \cofour\ and
\cione\ coverage is 0.2 dex smaller than the median value of the
galaxies that we consider for the remaining transitions. This might
imply an underestimate of the \lprimecofour\ and \lprimecione\
luminosities by the same factor, for constant $L'$/\lir\ ratios. 
However, this difference is well within the observed range of
ratios (V20) and it does not affect the essence of the results
presented in the coming sections. 

As in the previous sections, we did
not include AGN-dominated objects ($f_{\rm AGN}\geq80$\% of the total
\lir, Section \ref{sec:agn}) in the
analysis. A study of their SLEDs and the
contribution of X-ray dominated regions (XDR) is postponed to
future work. Nevertheless, we remark that the distribution of $f_{\rm
  AGN}$ is consistent between the
samples of main-sequence and the starburst objects analyzed in this Section.
For reference and completeness, we show the SLEDs for the
subsample with detected \cotwo\ and \cofive\ in Figure \ref{fig:app:sled} in Appendix \ref{app:cosled}.
We show the SLEDs in terms of
fluxes to facilitate the comparison with D15, where we normalized all
the curves to the mean \cotwo\ line flux of our main-sequence sample. 
The fluxes are computed at the median redshift of our sample
($z=1.25$), after averaging the luminosities to remove the
distance effect.
\begin{table}
  \centering
  \caption{Average emission line luminosities for galaxies at $z\sim1.25$.}
  \begin{tabular}{cccc}
    \toprule
    \toprule
    \multicolumn{4}{c}{Main-sequence}\\
    \midrule
    Transition & $N_{\rm det}, N_{\rm up}$& Mean& Median\tablefootmark{a}\\
    \midrule
    \lprimecotwo&    $18, 4$&  $1.83\pm0.23$\tablefootmark{$\dagger$}& $1.62^{+0.30}_{-0.70}$\\
    \lprimecofour&   $4, 0$  &  $0.66\pm0.08$& $0.71^{+0.16}_{-0.12}$\\
    \lprimecofive&    $20, 2$&  $0.52\pm0.06$\tablefootmark{$\dagger$}& $0.44^{+0.26}_{-0.11}$\\
    \lprimecoseven& $6, 0$  &  $0.17\pm0.04$& $0.17^{+0.01}_{-0.06}$\\
    \midrule
    \lprimecione&     $7, 1$  &  $0.37\pm0.05$\tablefootmark{$\dagger$}& $0.31^{+0.09}_{-0.07}$\\
    \lprimecitwo&     $6, 0$  &   $0.17\pm0.03$& $0.19^{+0.04}_{-0.11}$\\
    \midrule\\
    \multicolumn{4}{c}{Starbursts}\\
    \midrule
    Transition& $N_{\rm det}, N_{\rm up}$& Mean& Median\tablefootmark{a}\\
    \midrule
    \lprimecotwo&    $11, 1$& $1.91\pm0.24$\tablefootmark{$\dagger$}& $1.90^{+0.38}_{-1.06}$\\
    \lprimecofour&   $-$      & $-$& $-$\\
    \lprimecofive&    $15, 0$& $0.62\pm0.08$& $0.50^{+0.20}_{-0.10}$\\
    \lprimecoseven& $6, 0$  & $0.22\pm0.05$& $0.22^{+0.06}_{-0.10}$\\
    \midrule
    \lprimecione&     $-$      & $-$& $-$\\
    \lprimecitwo&     $6, 0$  & $0.20\pm0.04$& $0.19^{+0.05}_{-0.08}$\\
    \midrule\\
    \multicolumn{4}{c}{Extreme starbursts}\\
    \midrule
    Transition& $N_{\rm det}, N_{\rm up}$& Mean& Median\tablefootmark{a}\\
    \midrule
    \lprimecotwo&    $6, 1$& $1.80\pm0.27$\tablefootmark{$\dagger$}& $1.87^{+0.28}_{-1.05}$\\
    \lprimecofour&   $-$      & $-$& $-$\\
    \lprimecofive&    $6, 0$& $0.72\pm0.12$& $0.58^{+0.22}_{-0.13}$\\
    \lprimecoseven& $3, 0$  & $0.31\pm0.06$& $-$\\
    \midrule
    \lprimecione&     $-$      & $-$& $-$\\
    \lprimecitwo&     $3, 0$  & $0.27\pm0.06$& $-$\\
    \bottomrule
  \end{tabular}
  \tablefoot{  The $L'$ luminosities are expressed in $10^{10}$ \kkmspc. The
  average $I$ fluxes in \jykms\ shown in Figures \ref{fig:sled} and
  \ref{fig:lvg} are computed adopting $z=1.25$.\\
  The main sequence is parameterized as in
  \citet{sargent_2014}. Galaxies are defined as ``starbursts'' if
  $\Delta\mathrm{MS}\geq3.5$, and ``extreme starbursts'' if $\Delta\mathrm{MS}\geq7$.\\
  \tablefoottext{$\dagger$}{Formally biased mean value, as the first upper
  limit was turned into a detection for the calculation of the KM
  estimator \citep{kaplan-meier_1958}.}\\
  \tablefoottext{a}{The uncertainty is the interquartile range.}}
  \label{tab:sled}
\end{table}

\subsubsection{CO SLEDs across different galaxy populations}
\label{sec:sled:literature}
The average SLED for main-sequence galaxies at
$z\sim1.25$ appears significantly more excited than the disk of the Milky
Way, but not as excited as local ULIRGs, or high-redshift SMGs and
QSOs. Predictably, it is also substantially sub-thermally
excited already at mid-$J$ transitions
\citep{dannerbauer_2009}. On average, the $R_{52}=L'_{\rm
  CO(5-4)}/L'_{\rm CO(2-1)} = I_{\rm CO(5-4)}/I_{\rm CO(2-1)} /(J=5/J=2)^2$ ratio for
galaxies on the main sequence (Table \ref{tab:ratios}) is $1.8\times$ smaller than ULIRGs
\citep{papadopoulos_2012}, $1.3\times$ and $2.1\times$ than SMGs from
\cite{bothwell_2013} and \cite{spilker_2014}, and $2.4\times$ than
distant QSOs \citep{carilli_2013}. However, the $R_{52}$ ratio is $3.6\times$
higher than the observed values in the disk of the Milky Way
\citep{fixsen_1999}. Similar considerations apply for \coseven.

\subsubsection{CO SLEDs on and above the main sequence} 
\label{sec:sled:mainsequence}
\begin{table}
  \centering
  \caption{Average line luminosities ratios for galaxies at $z\sim1.25$.}
  \begin{tabular}{cccc}
    \toprule
    \toprule
    Transition & Main sequence& Starbursts& Extreme starbursts\\
    \midrule
    $R_{42}$&        $0.36\pm0.06$&  $-$& $-$\\
    $R_{52}$&        $0.28\pm0.05$&  $0.32\pm0.06$& $0.40\pm0.09$\\
    $R_{72}$&        $0.09\pm0.02$&  $0.12\pm0.03$& $0.17\pm0.04$\\
    $R_{\rm [CI]}$& $0.46\pm0.10$&  $-$& $-$\\
    \bottomrule
  \end{tabular}
  \tablefoot{The ratios and their uncertainties are computed
    analytically based on the mean $L'$
    luminosities in Table \ref{tab:sled}.
    Starbursts and extreme starbursts are defined as lying
    $\Delta\mathrm{MS}\geq3.5$ and $\geq7$ above the main sequence.\\
    $R_{42}=L'_{\rm CO(4-3)}/L'_{\rm CO(2-1)}$; $R_{52}=L'_{\rm
      CO(5-4)}/L'_{\rm CO(2-1)}$; \\
    $R_{72}=L'_{\rm CO(7-6)}/L'_{\rm
      CO(2-1)}$; $R_{\rm [CI]}=L'_{\rm [CI](2-1)}/L'_{\rm [CI](1-0)}$.}
  \label{tab:ratios}
\end{table}
The excitation of the average SLEDs only tentatively increases
with the distance from the main sequence. At mid/high-$J$ transitions, the $L'_J/L'_{\rm CO(2-1)}$ ratios are $1.1\times$ and $1.2\times$ higher
for starbursts than main-sequence galaxies for $J=5$ and $7$,
respectively (Table \ref{tab:ratios}). This difference and its low significance depend on the threshold for the
definition of starbursts, currently set at $\Delta\mathrm{MS}\geq3.5$; the
averaging of all galaxies in only two bins of $\Delta\mathrm{MS}$, further
softening the trend shown in Figure \ref{fig:distance}; and an intrinsic
diversity of shapes of CO SLEDs even within a sample of
homogeneously selected galaxies (Figure \ref{fig:app:sled}). We note
that the latter strongly affects any estimate of the molecular gas
mass from excited \co\ transitions.

A more
extreme threshold for the starburst regime results in an increase of the deviation
between the two samples and of its significance. The difference in $L'_J/L'_{\rm CO(2-1)}$ ratios rises to 
$1.4\times, 1.8\times$ at $J=5,7$ for $\Delta\mathrm{MS}\geq7$, 
substantially increasing the CO fluxes for the
starbursts and bringing them closer to the typical values of SMGs (Figure \ref{fig:sled}). The shape also looks
flatter, similarly to local IR-bright galaxies \citep{mashian_2015, kamenetzky_2016}.
However, this happens at the expense of the number
statistics, which are too sparse for a definitive conclusion about
such an extreme definition of starbursts.\\

The average \coratio\ ratio for
the IR-selected main-sequence objects is similar to the
previous estimate for the four $BzK$-selected galaxies from D15. We
note that the addition of
the $J=7$ transition constrains the peak
of the main-sequence SLED to lower $J$,
showing a significant departure from other more extreme populations of
galaxies. However, the overall shape of the SLED is flatter
than the rapid decrease observed in the disk of the Milky
Way, suggesting the existence of a secondary warm component and
excluding a steady increase at every $J$ observed so
far.

\subsubsection{Modeling of the CO SLEDs} 
\label{sec:sled:modeling}
The observed SLEDs allow for an assessment of the physically motivated
predictions from models and simulations. To simplify the comparison with previous
work, in Figure \ref{fig:sled} we show the same tracks reported in
D15: the empirical model from \cite{papadopoulos_2012} and the
hydrodynamical simulations from \cite{narayanan_2014} and
\cite{bournaud_2015}. \cite{papadopoulos_2012} assumes a hypothetical gas-rich disk
with a 10\% of the molecular gas in a star-forming phase with Orion A/B-like
excitation, along with a quiescent component with an
excitation as in the Milky Way. \cite{narayanan_2014} applied a 
radiative transfer code to simulated discs and mergers to calculate CO
SLEDs as a function of \sigmasfr. We
adopted their prescription for unresolved observations for galaxies with
$\Sigma_{\rm SFR}=1-10\,M_\odot\,\mathrm{yr^{-1}\,kpc^2}$, which are
typical for our main-sequence and starburst galaxies (Figure
\ref{fig:drivers}). \cite{bournaud_2015} applied a large
velocity gradient model to high-resolution simulations and compute a
synthetic CO emission. In particular, they distinguish the high
excitation of dense clumpy medium with the less
extreme conditions of the diffuse gas. Here we add the analytical
model for high-redshift star-forming galaxies by
\cite{vollmer_2017}. 
The simulations and the analytical model qualitatively reproduce
the rise of the CO SLED of main-sequence objects until mid-$J$
transitions and the following smooth decrease, while the constant
rising predicted by
\cite{papadopoulos_2012} does not appear to be followed by the average
observations.
The model by \cite{vollmer_2017} appears to best catch the
flat shape at high-$J$, while the simulations from
\cite{bournaud_2015} describe well the location of the peak for our
samples. The full treatment of the gas physics in simulations and the
analytical model seems to capture the main features of the SLEDs of
our sample, even if
the exact shape and the flux normalizations are
partially inconsistent with the observations. However, we warn the
reader that the shape of average observed SLEDs is influenced by
a mix of galaxies covering a range of excitation conditions, while the modeled
profiles are typical of each individual object (at least for the
simulations of \citealt{bournaud_2015}). 
For a definitive assessment of the various models, it will be
critical to extend the coverage to higher-$J$ emission, where their
predictions are mostly diverging (see \citealt{kamenetzky_2016} and
\citealt{vollmer_2017} for a discussion about
the performances for ULIRGs and SMGs at $J>6$).

\subsubsection{Large velocity gradient modeling} 
\label{sec:sled:lvg}
\begin{figure}
\includegraphics[width=\columnwidth]{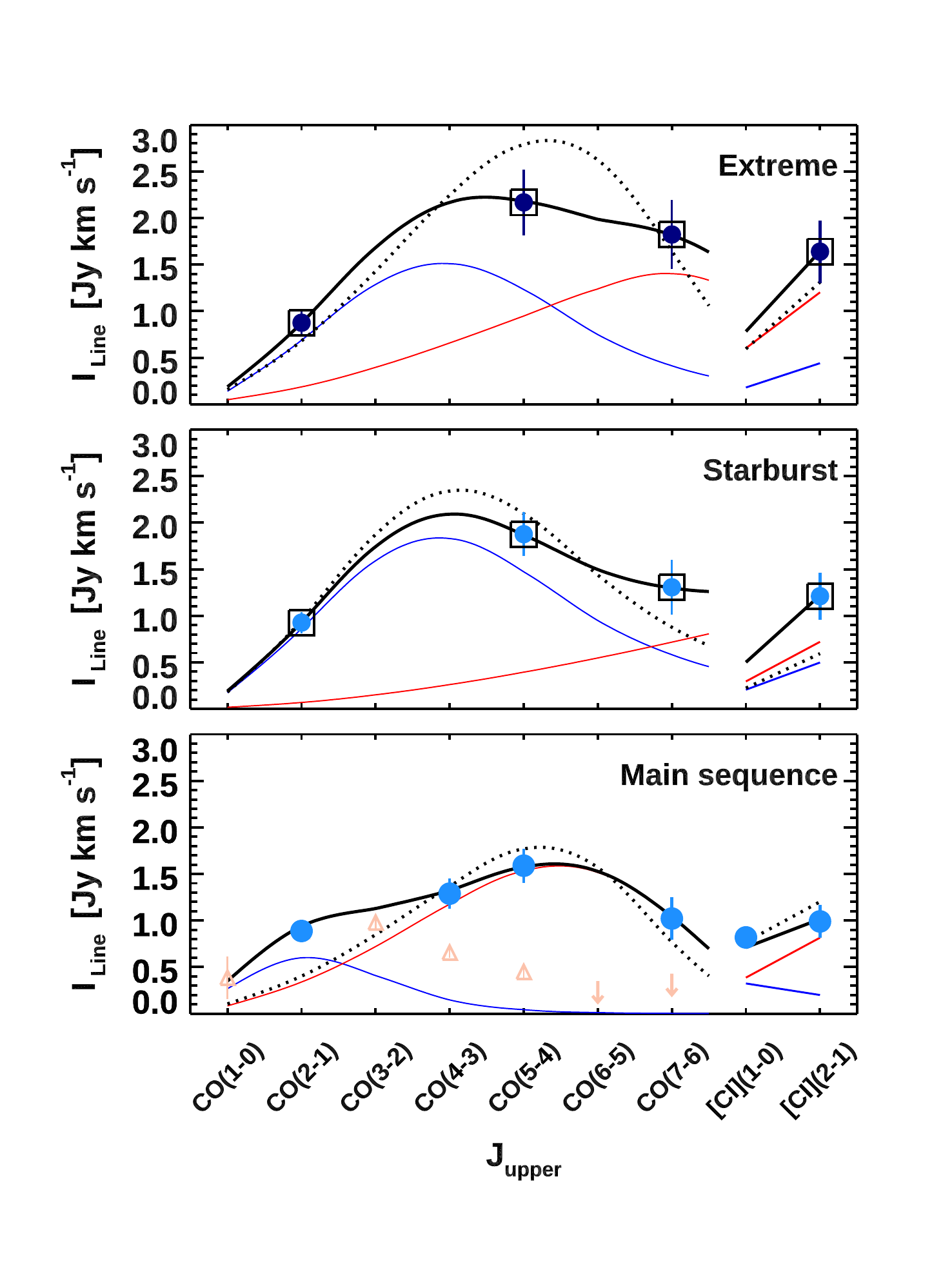}
\caption{\textbf{LVG modeling}. Large velocity gradient modeling of
  the observed CO+\ci\ SLEDs for main-sequence (\textit{bottom panel}),
  starburst ($\Delta\mathrm{MS}\geq 3.5$, \textit{central panel}), and
  extreme starburst galaxies ($\Delta\mathrm{MS}\geq 7$, \textit{top panel}) from our ALMA survey. The filled
  symbols show the mean fluxes. The blue and red lines show the
  low- and high-excitation components of the modeling, with the black solid line
  indicating their sum. The dotted black line shows the best-fit model
  with a single component. For reference, we show the CO
  data for the inner disk of the Milky Way \citep{fixsen_1999} normalized to the \cotwo\ emission of the average main-sequence
  galaxies as in Figure \ref{fig:sled} (pink open triangles). The upper limits are
  at $3\sigma$ significance. }
\label{fig:lvg}
\end{figure}
Large Velocity Gradient (LVG) modeling is a classical approach to gain
insight into the properties of the molecular gas in galaxies
\citep{goldreich_1974, scoville_1974, young_1991,papadopoulos_1998}. 
Here we followed the approach used in D15 (D. Liu et al. in prep.).
First, we used the RADEX tool \citep{vandertak_2007} to create a grid of
LVG models. We adopted the collisional rates from
\cite{flower_2001} with an ortho-to-para ratio of 3, and a CO
abundance to velocity gradient ratio of 
$[\mathrm{CO/H_2}]/(dv/dr)=10^{-5}\,\mathrm{km\,s^{-1}pc^{-1}}$ valid
for solar metallicities \citep{weiss_2005_sled, weiss_2007}.
We computed a
model grid for the median redshift of the sample ($z=1.25$), covering density and
temperature intervals of $n(\mathrm{H}_2)=10^2-10^6\,\mathrm{cm}^{-3}$ and $T_{\rm kin}=
5-300$ K, including the appropriate value of the temperature of the
cosmic microwave background. 
We fixed the line width to $50$ \kms\ or, equivalently, the cloud scale height 
to 10 pc, values typical of giant molecular clouds (GMCs). Given the limited amount
of information, leaving these parameters or $dv/dr$ free to vary would result in
overfitting. Considering that a galaxy contains many
of these LVG clouds, the beam filling factor is simply their number. 
We caution the reader that due to the high degeneracy among the LVG
parameters, even with a handful of CO lines one
could not obtain results simultaneously constraining $n_{\rm H_2}$,
$T_{\rm kin}$, and $N(\mathrm{H_2})$. Nevertheless, we
verified that our estimates of $N(\mathrm{H_2})$ and the line optical depths are within
the reasonable physical ranges for GMCs
\citep{glover_2015,tress_2020}. All things considered, the best-fit $n_{\rm H_2}$
and $T_{\rm kin}$ in this work mostly reflect the relative trends
between the different subsamples of main-sequence and starburst galaxies.\\

We determined the best-fit model via a
customized $\chi^2$ minimization algorithm, optimized for the exploration
of highly multi-dimensional spaces
(\textsc{MICHI2}\footnote{\url{https://ascl.net/code/v/2533}}, \citealt{liu_2020_michi2}).
In particular, we iteratively sampled the $\chi^2$ distribution $15,000$ times,
randomizing the parameters within normal distributions centered on the
lowest $\chi^2$ derived at the previous iteration, but artificially inflating their width.
The output consists in best-fit (i.e., $\mathrm{min}(\,\chi^2)$)
parameters and their $\sigma$ uncertainties, plus median and $68\%$
interpercentile values for an arbitrarily large number of components. 
To better constrain the fit, we further included the \ci\ transitions
under the assumption that the neutral atomic carbon is co-spatial with CO. We generated the
models with RADEX assuming a fixed abundance of
$\mathrm{[CI]/H_2} = 3\times10^{-5}$ \citep[but see V18 on the
reliability of this assumption]{weiss_2003,
  papadopoulos_2004}. We independently fitted the line fluxes for the average
main-sequence and starburst samples (Table \ref{tab:sled}).\\ 

It is evident from Figure
\ref{fig:lvg} that single components do not provide a good representation to the
observed CO SLEDs of both populations. This is was already suggested
for $BzK$ galaxies (D15, see also \citealt{brisbin_2019}), and it is a well known fact for local IR-bright
galaxies \citep{papadopoulos_2010_sled, papadopoulos_2012,
  lu_2014, rosenberg_2015, kamenetzky_2016} and high-redshift SMGs and QSOs
\citep[e.g.,][]{weiss_2007, aravena_2008, ivison_2011, bothwell_2013, carilli_2013, greve_2014,
  spilker_2014, yang_2017, canameras_2018}. This result still holds
when modeling only the \co\ emission, excluding the \ci\ transitions. \\

The addition of a second
component outperforms the previous attempt.
We assumed the existence of a dense and a diffuse phase
by imposing that $n_{\rm H_2,\,low}<n_{\rm H_2,\,high}$. This
results in better constrained densities, but not \tkin\ (Table \ref{tab:lvg}). This
is particularly evident for the starbursts, likely due to the lack of
mid-$J$ coverage.
For both main sequence and starburst galaxies we retrieve the existence of
similar low- ($n_{\rm H_2,\,low}\sim10^{2}-10^{3}\,\mathrm{cm}^{-3}$) and
high-density components  ($n_{\rm H_2,\,high}\sim10^{4}-10^{6}\,\mathrm{cm}^{-3}$).
Furthermore, we retrieve a substantial amount of gas in the
dense phase. The latter encloses $\sim50$\% of the total
molecular gas mass for main-sequence galaxies. The fraction of denser gas in
starbursts is hardly constrained at this stage, possibly due to the
absence of lines at $J>7$, where a substantial emission from the
excited component might
be expected, as shown in local ULIRGs \citep[e.g.,][]{
  mashian_2015, kamenetzky_2016} and distant SMGs \citep[e.g.,][]{yang_2017, canameras_2018}.  
This is suggested by the flatter shape of the SLEDs for the more
extreme starburst with $\Delta\mathrm{MS}\geq 7$.
We note that the absolute values of the gas mass from this modeling depend on the adopted CO
abundance, constant for both main sequence and starburst
galaxies. Therefore, they are subject to the uncertainties already
described in Section \ref{sec:drivers:umean}. Relative comparisons
between the two phases for each population still hold, under the
assumption that dense and diffuse gas reservoirs share the same metallicity.
\begin{table}
  \centering
  \caption{Best-fit parameters of a double-component LVG modeling of
    the observed CO+\ci\ SLEDs of galaxies at $z\sim1.25$.}
  \begin{tabular}{ccc}
   \toprule
   \toprule
   \multicolumn{3}{c}{Main-sequence}\\
   \midrule
   & Low& High\\
   \midrule
   $\mathrm{log}\left( n_{\rm H_2} / [\mathrm{cm^{-3}}] \right)$ & $2.2\pm0.3$ & $3.9\pm0.1$\\
   $T_{\rm kin} / [\mathrm{K}]$&  $45\pm113$& $45\pm5$  \\
   \midrule\\
   \multicolumn{3}{c}{Starburst}\\
   \midrule
   & Low& High\\
   \midrule
   $\mathrm{log}\left( n_{\rm H_2} / [\mathrm{cm^{-3}}] \right)$ &  $2.9\pm0.8$ & $6.1\pm1.4$\\
   $T_{\rm kin} / [\mathrm{K}]$&  $300\pm138$& $75\pm138$  \\
   \midrule\\
   \multicolumn{3}{c}{Extreme starburst}\\
   \midrule
   & Low& High\\
   \midrule
   $\mathrm{log}\left( n_{\rm H_2} / [\mathrm{cm^{-3}}] \right)$ &  $3.0\pm0.8$ & $4.5\pm1.3$\\
   $T_{\rm kin} / [\mathrm{K}]$&  $215\pm138$& $35\pm138$  \\
   \bottomrule
 \end{tabular}
 \tablefoot{The average values and their uncertainties are the
   best-fit estimates and their statistical errors, where we imposed
   that $n_{\rm H_2,\,low}<n_{\rm H_2,\,high}$.} 
 \label{tab:lvg}
\end{table}

\subsubsection{The effect of dust opacity on the high-$J$ CO emission}
\label{sec:opacity}
Large dust optical depths even at (sub-)mm wavelengths are responsible
for the apparent depressed high-$J$ CO emission in extreme objects
as Arp 220 \citep{greve_2009, papadopoulos_2010_arp220, rangwala_2011,
  scoville_2017_arp220}. This is due to the fact that for $\tau_{\rm
  dust}\gg 1$, the line emissions are largely erased by the quasi-black body dust
continuum \citep{papadopoulos_2010_arp220}. A modeling of the dust
continuum emission leaving the optical depth free to vary can
provide meaningful results only when the SED is well sampled from
mid-IR to mm wavelengths, and it might still return degenerate
solutions with the temperature \tdust, in absence of independent ways to
distinguish them \citep{cortzen_2020}. Indeed, our SED modeling is based on
the assumption that the dust emission is optically thin above
$100\,\mu\mathrm{m}$.

Here we tested our assumption by computing the dust optical depth at the \cofive\ and \coseven\
wavelengths as $\tau_{\rm dust} = \kappa(\nu)\,\Sigma_{d} = \kappa_{\rm
  850\,\mu m}\,( \nu/\nu_{\rm 850\,\mu m})^{\,\beta}\,M_{\rm
  dust}/(2\pi R^2)$, where $\kappa$ is the
frequency-dependent dust opacity, and $\Sigma_{d}$ the dust mass
surface density \citep[e.g.,][]{casey_2014}. We adopted $\kappa_{\rm
  850\,\mu m}=0.43\,\mathrm{cm^{-2}g^{-1}}$ and $\beta=2$
\citep{li_2001}, the \mdust\ from the SED modeling (Section
\ref{sec:sed_modeling}), and the sizes from the ALMA measurements. We
note that, while using one of the outputs of the SED fitting, this
sanity check is not tautological, given the introduction of the size
in the calculation.
For sources with a significant \cofive\ line detection, a determination of
\mdust\ and of the size from ALMA, we retrieve $\langle \tau_{\rm
  dust}\,(520\mu\mathrm{m})_{\rm CO(5-4)} \rangle=0.012\pm0.003$, including the lower
limits on $\tau_{\rm dust}$ due to the upper limits on the sizes. For the dust
continuum emission under \coseven, we find $\langle \tau_{\rm
  dust}\,(371\mu\mathrm{m})_{\rm CO(7-6)} \rangle=0.020\pm0.007$, with a maximum of
$0.08$, where all the \coseven\ detections have a safe determination of
their size. We note that the same calculation with \mdust\ derived from an
ideal SED modeling with the opacity as a free parameter would be
lower, further decreasing the value of $\tau_{\rm dust}$. \\

The largest opacities are associated with
strongly starbursting galaxies and/or AGN contamination, but their
observed high-$J$/low-$J$ CO ratios do not appear systematically
depressed compared to the rest of the sample. However, they do have
$\tau_{\rm dust}=1$ for rest-frame $\lambda\sim100-140$ $\mu$m, similar to
several high-redshift SMGs and indicating that their \mdust\ are likely
overestimated and \tdust\ cooler than what they really are
\citep{jin_2019, cortzen_2020}. These cases represent $<5$\% of the sample
for which we could carry out the test on the opacity, and they do not
influence the final results. Therefore, it
appears that the dust opacity does not have a significant impact on
the emission of mid- and high-$J$ CO lines on galaxy scales in our
sample, and this is due to the less extreme $\Sigma_{\rm dust}$ compared
to, e.g., the one of Arp 220. However, we notice that this is a point
to be reassessed with higher spatial resolution measurements, which
might well reveal compact pockets of gas within our targets more
affected by the dust absorption.

\section{Discussion}
\label{sec:discussion}
In the previous sections, we showed that, on global scales, the CO
line emission and excitation of main sequence and starburst galaxies
broadly correlate with a variety of properties.
The high-$J$ CO line luminosities ($J=5, 7$) are
quasi-linearly related to the star formation rate, suggesting a
physical connection with the gas pockets where new stars are
formed. The low-$J$ CO line emission is associated with less dense
molecular gas, tracing the bulk of its mass in galaxies.
Interestingly, the \lprimecofive/\lprimecotwo\ ratio and overall CO
SLED increase as a function of the total infrared luminosity \lir\
($\propto\mathrm{SFR}$), the mean intensity of the radiation field heating up
the dust \umean\ ($\propto T_{\rm dust}$), the star formation
efficiency $\mathrm{SFE}=\mathrm{SFR}/M_\star$, the density of the
  SFR (\sigmasfr), and, less distinctly, with the distance from the main 
sequence \distms.
A comparison of the strength of the observed correlations and their intrinsic
scatter (Table \ref{tab:scalingrelations}) offers further insight into
this network of properties.\\

\subsection{The spatial distribution of SFR as the driver of
  the properties of star-forming galaxies}
The main physical driver of the CO excitation seems to be a combination
of the amount of star formation occurring in the galaxy and its
spatial distribution. While a SFR-$R_{52}$ correlation does exist, its
strength increases by using \sigmasfr, instead.  
This naturally follows the fact that dense gas concentrations ignite
more compact star-forming regions, producing large UV radiation
fields and cosmic ray rates, and warming up the dust \citep{narayanan_2014}. This is reflected on the
similarly strong and tight \umean-$R_{52}$ relation, and on the
enhancement of the excitation as a function of the SFE, boosted in 
more compact gas configurations \citep{papadopoulos_2012}. Note that
the correlation between the \co\ excitation and \umean\ does not
necessarily imply that the interstellar radiation field is responsible
for the excitation of the mid-/high-$J$ transitions. If multi-component
PDRs have been shown to be sufficient to describe the CO SLED of local
spirals \citep{rigopoulou_2013}, this is not the case for starbursts
where mechanical heating induced by SF-related (supernovae) or
unrelated (mergers, AGN outflows, radio-jets) shocks are invoked to explain the observations
(\citealt{rangwala_2011, kamenetzky_2012, lu_2014, kamenetzky_2016};
see \citealt{brisbin_2019} for the case of a distant
main-sequence galaxy). The correlation
with \umean\ would then be indirect: it is the stellar feedback
from the intense star formation to drive it \citep{wu_2015}. This
might well be the case for the most extreme starbursts, for which PDRs
cannot reproduce the observed flat CO SLEDs \citep{kamenetzky_2016}. We
note that the models in Figure \ref{fig:sled} specifically contemplating a recipe for the
mechanical heating from SNae better reproduce the flat shapes of the
CO SLEDs and the location of their peaks (\citealt{bournaud_2015, vollmer_2017}, and see the discussion
in \citealt{papadopoulos_2012}).\\

The global rise of $R_{52}$ with the distance from the main sequence can be
interpreted considering that
\sigmasfr\ overall increases with \distms\
\citep{elbaz_2011}. Moreover, the high gas
fractions inducing clumping in turbulent
high-redshift massive disks further enhance this effect
\citep{bournaud_2015}, increasing
the level of CO excitation of distant main-sequence galaxies with
respect to local Milky Way-like objects. In other words, the
same mechanisms that we consider here as acting on global galaxy
scales might well be in action in sub-galactic massive clumps, effectively
mimicking starburst environments
\citep{zanella_2015,zanella_2019}. This is also consistent
with the results from LVG modeling, where the density of the
collisionally excited gas is the term driving the
high-$J$ emission, given the fast de-excitation rates. 
Finally, taken to its extreme
consequences, the formation of hundreds or even thousands of stars per
kpc$^2$ and the ensuing massive layers of dust surrounding them
\citep[e.g., Arp 220,][]{barcos_munoz_2015, scoville_2017_arp220}
affects the emerging CO spectrum of a galaxy, inducing the uttermost 
opacities at long wavelengths \citep[e.g.,][]{blain_2003, greve_2009, 
  riechers_2013, huang_2014,
  lutz_2016, hodge_2016, spilker_2016, simpson_2017}, reducing the
apparent CO excitation at high-$J$
\citep{papadopoulos_2010_arp220,rangwala_2011} and the dust temperature,
affecting the dust mass estimates if not properly taken into account
\citep{jin_2019, cortzen_2020}.

\subsection{What is a starburst?}
The scenario presented above has been formulated in various flavors to
individually explain several of the properties reported here. The main
addition of this work, namely the excitation of CO in distant main-sequence
and starburst galaxies, fits in the
general picture that we sketched. The ensemble of
properties and correlations we reported here can also be used to revisit
the definition of what a ``starburst'' is. A standard operational
classification is based on the distance from the observed
empirical \mstar-SFR correlation, the main sequence. This proved to be
a useful distinction and an excellent predictor of several trends
\citep[e.g.,][]{sargent_2014}, but recent results, including
our present and previous analysis \citep{puglisi_2019}, show that the
demarcation between starburst and main sequence galaxies is more
blurred that we previously considered. We do find ``starburst-like''
behaviors on the main sequence \citep{elbaz_2018}, likely linked to
the existence of transitional objects \citep[][A. Puglisi et al. in
prep., to limit the references to recent works based on sub-mm
observations]{popping_2017, barro_2017_co,
  gomez-guijarro_2019b, puglisi_2019}. Such transition might well
imply an imminent increase of the SFR, driving the object in the realm
of starbursts \citep[e.g.,][]{barro_2017_co}, or its cessation,
bringing the system back onto or even below the main sequence
\citep{gomez-guijarro_2019b, puglisi_2019}, with the CO properties
potentially able to distinguish between these two
scenarios. Regardless of these transitional objects, a definition of
starburst based on \sigmasfr, rather than \distms, would naturally
better account for the observed molecular gas excitation properties,
dust temperatures and opacities, or SFE \citep[see also][]{elbaz_2011,
  rujopakarn_2011, jimenez-andrade_2018,
  tacconi_2020}. As an example, in Figure
\ref{fig:sled} we show the mean SLED of the subsample of galaxies with
both \cotwo\ and \cofive\ coverage, split at its median
\sigmasfr. While only tentative at this stage, this suggests a
trend of increasing CO excitation with \sigmasfr, consistently with
Figure \ref{fig:drivers} and what mentioned above.\\

In more physical terms, the new definition
would trace the observed correlations back to a common origin: the
accumulation of gas and formation of stars in compact configurations,
following global or local dynamical changes in the galaxy structure.
The latter might be due to major mergers, known to be primary drivers of
starbursts activity in the local Universe
\citep{sanders_mirabel_1996}, or in presence of higher gas fractions,
to minor mergers \citep{bustamante_2018, gomez-guijarro_2018} or violent disk instabilities induced by a sudden alteration of the
gravitational equilibrium, particularly effective at high redshift
\citep[e.g.,][]{bournaud_2007, ceverino_2010, dekel_2014}. Such gas
concentrations would further increase the AGN activity
\citep{elbaz_2018}, spread also in our sample. Moreover, a definition based on
\sigmasfr\ would allow for the classification of SMGs, normally hard
to achieve because of the lack of \mstar\ determinations, potentially 
sorting extended massive disks \citep{hodge_2016, hodge_2019}, and
bona fide ongoing gas-rich mergers \citep[see][for a review of current
models for the formation of dusty star-forming galaxies]{casey_2014}.
We will explore the detailed properties of our sample as a function of 
size and compactness in a dedicated work (A. Puglisi et al. in prep.).

\section{Conclusions}
We presented the outcome of a multi-cycle ALMA survey of the CO emission in
IR-selected galaxies on and above the main sequence at
$z\sim1.3$. We obtained new observations of low- to high-$J$
lines that we complemented with existing samples of local and distant
star-forming galaxies. In detail:
\begin{enumerate}
  \item We report new detections of \cofive, \cotwo, and
   \coseven(+\citwo) for 50, 33, and 13 galaxies, respectively,
   corresponding to detection rates of $\sim50-80$\%. 
  \item We found that the \cofive\ and \coseven\ luminosities of both
    main-sequence and starburst galaxies follow 
    an almost linear and tight ($\sigma_{\rm int}=0.16$ dex) correlation
    with the total \lir, as previously established for local IR-bright
    galaxies, spirals, and distant SMGs. On the other hand, the
    \cotwo\ emission of main sequence and starburst is consistent with
    the
    integrated SK law. This
    suggests that the \cofive\ and \coseven\ emission is associated with
    the reservoirs of actively star-forming gas in galaxies, while
    \cotwo\ traces the total mass of cold and less dense molecular
    medium. This also
    suggests caution when deriving the total molecular \mgas\ from high-$J$ CO
    transitions. 
  \item Moreover, we found the \cotwo/\lir\ ratio to steadily decrease as a
    function of the distance from the main sequence \distms, while
    \cofive/\lir\ and \coseven/\lir\ remain constant. This
    further supports the idea that mid- and high-$J$
    transitions trace the SFR, independently of
    their stellar mass and redshift, within the parameter space
    spanned by our observations.
  \item We derived monotonically increasing \cofive/\cotwo\ luminosity
    ratios -- a proxy for the CO excitation -- as a function of
    increasing star formation efficiency, mean
    intensity radiation field \umean, SFR surface density
    \sigmasfr, and, less distinctly, \distms. 
  \item We found the overall CO SLED of distant main-sequence galaxies up to
    \coseven\ to be more excited than the disk of the Milky Way, but
    less than local ULIRGs or
    high-redshift SMGs and QSOs. An intrinsic variety of shapes is present, as shown by
    the dispersion of the observed CO luminosity ratios, blurring the
    distinction between the SLEDs of starbursts and
    upper main-sequence objects. 
    The dust opacity does not appear to
    significantly suppress the high-$J$ CO emission even for the most
    extreme objects in our sample, due to relatively low dust mass
    surface densities compared to, e.g., Arp 220, the prototypical
    case for this matter. However, this has to be further tested with
    observations at higher spatial resolution. 
  \item We modeled the observed CO(+\ci) SLEDs adopting the LVG
    method. The addition of high-$J$ CO and \ci\ lines indicates the
    existence of a second highly excited component both for starbursts
    and main-sequence galaxies, similarly to what
    invoked to explain the SLEDs of local ULIRGs and SMGs. Imposing
    the existence of a dense and a diffuse component, we retrieve substantial amount of
    gas in the former phase ($n_{\rm H_2,\,high}\sim10^{4}-10^{5}$ cm$^{-3}$),
    contributing to $\sim50$\%
    of the total molecular gas mass for main-sequence galaxies.
  \item We interpret the CO excitation conditions as driven by the
    combination of large SFRs over compact regions. Such large
    \sigmasfr\ values naturally explain the large gas densities and
    high temperatures due to increased UV radiation fields, cosmic ray
    heating, and dust and gas coupling. Larger densities also naturally
    induce enhanced SFEs, as canonically advocated for starbursts.
    An operational definition based on
    \sigmasfr\ rather than on the offset from the main sequence might
    better separate truly
    starbursting galaxies from secularly evolving disks.
  \item Idealized simulations, analytical, and semi-empirical models
    qualitatively account for the increase of the CO excitation in distant
    main-sequence and starburst galaxies peaking around $J\sim4-5$, but starting from different premises and resulting
    in shapes and normalizations partially inconsistent with the average observed
    trends. The addition of transitions at $J>7$ will be the key for a
    definitive assessments of several models.

\end{enumerate}

\begin{acknowledgements}
  We acknowledge the constructive comments from the anonymous referee
  that improved the content and presentation of the results. 
  We thank Jonathan Freundlich for his guidance in using the
  PHIBSS-2 data products. F.V. acknowledges support
  from the Carlsberg Foundation Research Grant CF18-0388
  ``Galaxies: Rise and Death''. F.V. and G.E.M. acknowledge the Villum
  Fonden Research Grant 13160 ``Gas to stars, stars to dust:
  tracing star formation across cosmic time'' and the Cosmic Dawn
  Center of Excellence funded by the Danish National Research
  Foundation under then Grant No. 140. G.E.M. acknowledges support from the
  European Research Council (ERC) Consolidator Grant funding
  scheme (Project ConTExt, Grant No. 648179). D.L. acknowledges
  funding from the European Research Council (ERC) under the
  European Union's Horizon 2020 research and innovation programme
  (grant agreement No. 694343). Este trabajo cont\'o 
  con el apoyo de ``CONICYT+PCI+INSTITUTO MAX PLANCK DE
  ASTRONOMIA MPG190030''. M.A. has been supported by the grant
  ``CONICYT+PCI+REDES 190194''. H.D. acknowledges financial support
  from the Spanish Ministry of Science, Innovation and Universities
  (MICIU) under the 2014 Ram\'{o}n y Cajal program RYC-2014-15686 and
  AYA2017-84061-P, the later one co-financed by FEDER (European
  Regional Development Funds). 
  S.J. acknowledges financial support from the Spanish Ministry of
  Science, Innovation and Universities (MICIU) under grant
  AYA2017-84061-P, co-financed by FEDER (European Regional Development
  Funds). Y.G.'s research is supported by  
  National Key Basic Research and Development Program of China (grant 
  No. 2017YFA0402704),  
  National Natural Science Foundation of China (grant Nos. 11861131007,
  11420101002), and Chinese Academy of Sciences Key
  Research Program of Frontier Sciences (grant No. QYZDJSSW-SLH008).
  In this work
  we made use of the COSMOS master spectroscopic catalog --
  kept updated by Mara Salvato --, of GILDAS, and STSDAS. GILDAS,
  the Grenoble Image and
  Line Data Analysis Software, is a joint effort of IRAM and the
  Observatoire de Grenoble. STSDAS is a product of the Space Telescope
  Science Institute, which is operated by AURA for NASA. Moreover,
  this paper makes use of the following ALMA data: ADS/JAO.ALMA,
  \#2019.1.01702.S, \#2018.1.00635.S, \#2016.1.01040.S, \#2016.1.00171.S, and
  \#2015.1.00260.S. ALMA is a partnership of ESO (representing its
  member states), NSF (USA) and NINS (Japan), together with NRC
  (Canada), MOST and ASIAA (Taiwan), and KASI (Republic of Korea),
  in cooperation with the Republic of Chile. The Joint ALMA
  Observatory is operated by ESO, AUI/NRAO, and NAOJ. 
\end{acknowledgements}

\bibliography{bib_cosurvey} 
\bibliographystyle{aa}

\appendix
\section{Total recovered fluxes}
\label{app:totalflux}
The total flux of a source can be robustly recovered if its size is
securely estimated. However, our iterative extraction could drive to
flux losses when we estimate only an upper limit on the size,
and such value is comparable with the beam size. We estimated these
losses as detailed in Appendix B of V20, namely by
injecting artificial bright galaxies with circular Gaussian
profiles and a FWHM fixed to the $1\sigma$ upper limit on the size in the
\textit{uv} plane, and then re-extracting their fluxes with the fiducial point
source profile. We then corrected the extracted fluxes to ($I_{\rm
  Gauss}/I_{\rm Point} + 1)/2$ and added in quadrature the absolute error on
such correction ($\sigma_{\rm corr} = (I_{\rm Gauss} - I_{\rm
  Point})/2$) to the statistical uncertainty. The correction does not
depend on the brightness of the injected mock source, provided that it is
significantly detected ($\mathrm{SN}\geq10$), nor on its position
in the map. The sizes and the flux corrections are reported in the
data release.

\section{Revisiting the target selection a posteriori}
\label{app:selection}
\begin{figure}
\includegraphics[width=\columnwidth]{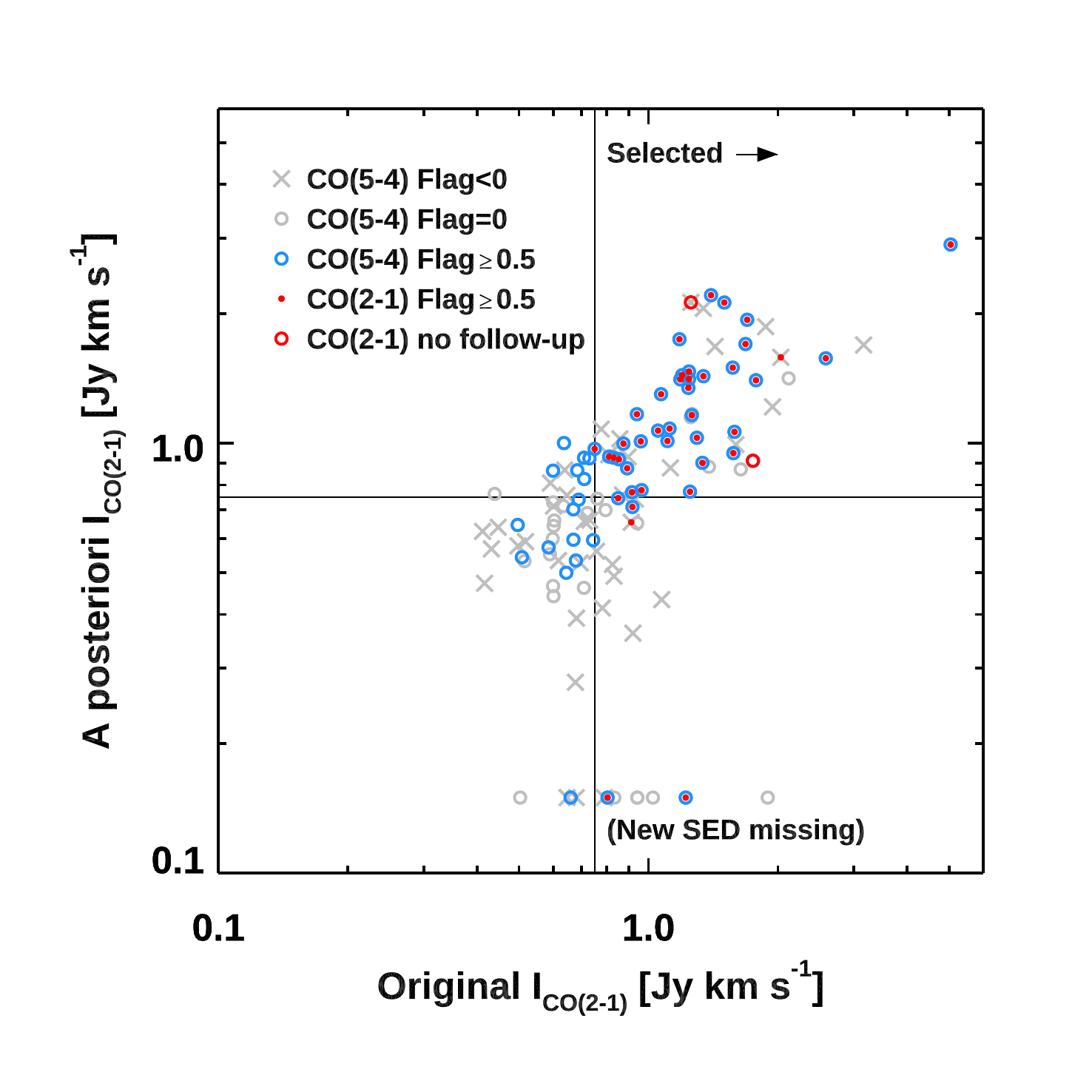}
\caption{\textbf{Revisiting the selection and the impact on the
    detection rate}. Predicted \icotwo\ fluxes with the original
  \lir\ based on PACS detections and on the new IR photometry from
  \cite{jin_2018}. The symbols mark the whole parent sample of 123
  galaxies with ALMA Band 6 observations targeting \cofive. Gray
  crosses, gray open circles, and blue open circles indicate sources with
  uncertain and reliable information on \cofive\
  ($Flag<0,\,0,\,\geq0.5$, respectively, Section \ref{sec:flag_success}).
  Red filled circles show galaxies with
  reliable information on \cotwo\ ($Flag\geq0.5$). Open red circles mark objects with
  predicted fluxes bright enough to be observed according to both the
  initial and the revised SED modeling, but that did not
  enter the final selection due to the frequency grouping. 
  The black solid lines indicate the
  depth of the ALMA Band 3 observations: objects on the right side of
  the vertical line were selected for the follow-up (excluding the red
  open circles); object above the
  horizontal line could have been selected, if the updated
  photometry were available when preparing the observations. 
  Objects with missing photometry in \cite{jin_2018} are
  artificially set to $I_{\rm CO(2-1)}=0.15$ \jykms\ and labeled accordingly.}
\label{fig:selection}
\end{figure} 
Reconstructing a posteriori the target selection for the \cotwo\
follow-up observations, with the improved constraints on the IR
photometry that became available in the meantime, allows us to get a handle on
the factors determining the detection rates described in Section
\ref{sec:flag_success}. In Figure \ref{fig:selection}, we show the original
prediction of the $I_{\rm CO(2-1)}^{\rm pred}$ fluxes computed from previous \lir\ estimates 
from the PEP survey (Section \ref{sec:sample}), against an updated version based on
the far-IR modeling of the deblended photometry from
\cite{jin_2018}. Excluding 12 sources from the PEP survey without a
counterpart in the deblended catalog, the flux predictions scatter
around the one-to-one relation. Galaxies in the bottom right quadrant
of Figure \ref{fig:selection} were bright enough to be selected for
the follow-up observations, but they
would have missed the cut based on the updated \lir. The negligible
fraction of \cotwo\ detections among these objects supports the
hypothesis that they are indeed too faint to be detected at the current
depth. On the other hand, for the sources in the top right quadrant of
Figure \ref{fig:selection}, the new photometric modeling supports the
initial selection. In fact, this is where 
virtually all \cotwo\ detections are located. Undetected
objects primarily lack reliable information of \cofive\ and have
low quality flags on the optical/near-IR spectroscopic redshifts
\zopt. Only a minor fraction of IR-bright sources do have secure
\zopt, but
remained undetected in \cofive\ and \cotwo, likely due to bona fide
dimmer line fluxes than predicted. Galaxies in the remaining left
quadrants were not selected for the \cotwo\ follow-up and their
\cofive\ detection rate is set by a combination of bad \zopt, lower
\lir\ than previously estimated, and intrinsically faint lines in
bright objects, in order of importance. The latter are physically
interesting, but we cannot currently put any constraints on their
properties, in absence of a secure \zsubmm. 

\section{The diversity of the CO SLEDs of galaxies on and above the main sequence}
\label{app:cosled}
\begin{figure}
\includegraphics[width=\columnwidth]{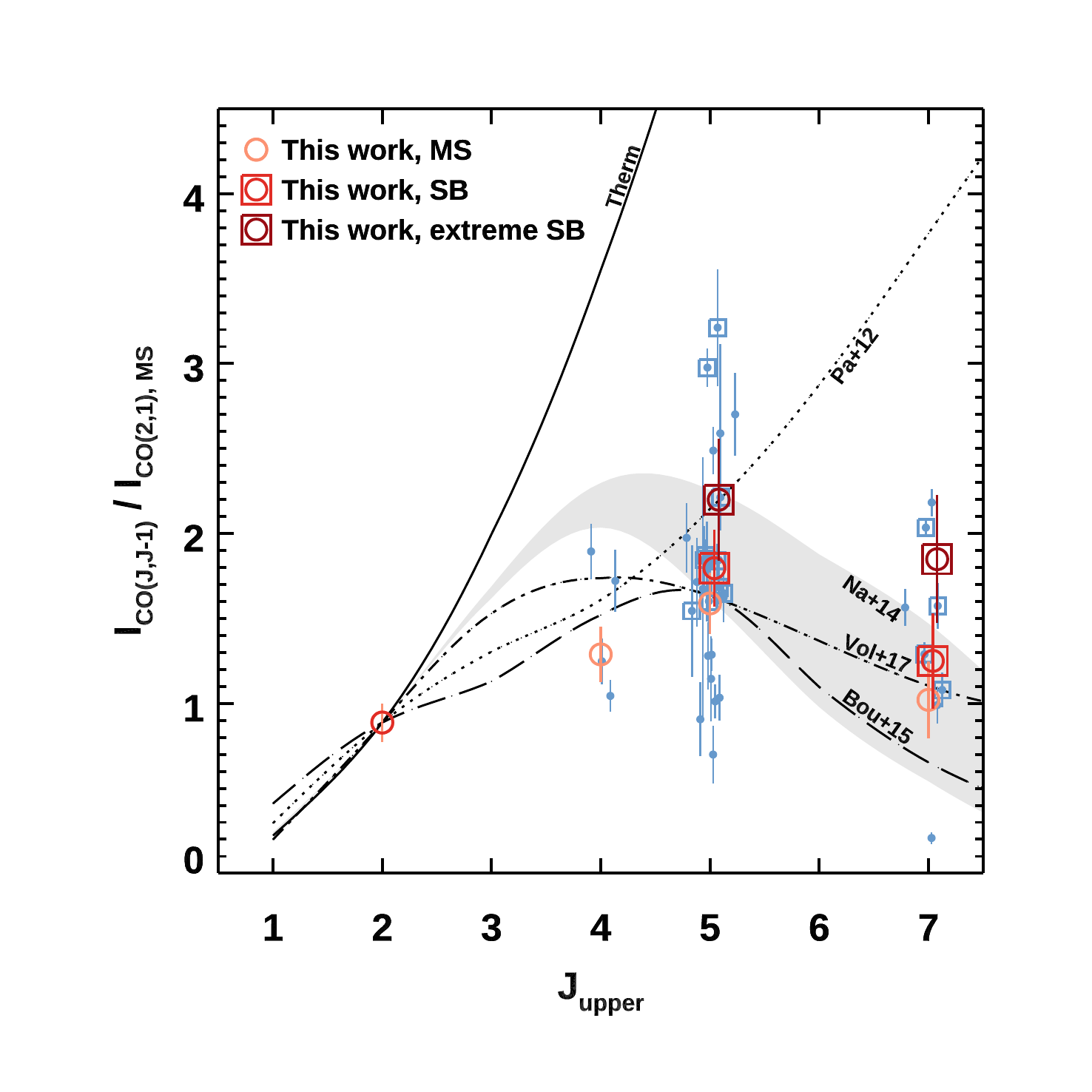}
\caption{\textbf{Diversity of CO SLEDs for main-sequence and
    starburst galaxies.} Blue filled circles and open squares
  indicate the main-sequence and starburst galaxies with detected
  \cofive\ and \cotwo\ lines from our ALMA follow-up, normalized to
  the mean \cotwo\ flux for the main-sequence sample. The open red
  circles and squares indicate the mean fluxes for main-sequence,
  starburst ($\Delta\mathrm{MS}\geq 3.5$), and extreme starburst
  ($\Delta\mathrm{MS}\geq 7$) objects. The solid line
  shows the line ratios for a fully thermalized case. The gray shaded
  area marks the model by
  \cite{narayanan_2014} for unresolved
  observations within $\Sigma_{\rm SFR} = 1-10\,M_\odot\,\mathrm{yr^{-1}\,kpc^2}$. The 
  long dashed line indicates the simulations from
  \cite{bournaud_2015}. The dotted line tracks the empirical model by
  \cite{papadopoulos_2012}. The dashed-dotted line points at the
  analytical model by \cite{vollmer_2017}.}
\label{fig:app:sled}
\end{figure}  
In Figure \ref{fig:app:sled} we show the CO SLEDs for the subsample of
main-sequence and starburst galaxies with \cotwo\ and \cofive\
detections from our ALMA observations, normalized to the mean \cotwo\
flux for the main-sequence sample (Section \ref{sec:cosled}). This
figure highlights the variety of shapes displayed even by a
homogeneously selected sample of normal and extreme galaxies at high redshift.

\section{Data tables, galaxy spectra and spectral energy distribution}
\label{app:spectra_seds}
Figure \ref{fig:spectra} shows an example of the ALMA spectra and the IR SED for our
sample of reliable sources used in the analysis ($Flag \geq
0.5$). The whole compilation of spectra from which we extracted
reliable information is available in the online version of the
article. Similarly, the full data table is made public in .fits
format. The description of the columns is listed in Table \ref{tab:datatable}. 
\begin{table*}
\small
  \centering
  \caption{Column description for the data release.}
  \begin{tabular}{lcl}
    \toprule
    \toprule
    Name & Units & Description\\
    \midrule
    ID& \dots& Identifier\\
    R.A.& hh:mm:ss& Right ascension\\
    Dec&  dd:mm:ss& Declination\\
    zspec\textunderscore opticalnir& \dots& Optical/near-IR
                                            spectroscopic redshift
                                            (M. Salvato et al., in prep.)\\ 
    (d)zspec\textunderscore submm& \dots& ALMA sub-mm spectroscopic redshift\\ 
    Log\textunderscore StellarMass& \msun& Logarithm of the stellar mass (\citealt{chabrier_2003} IMF)\\
    (d)Total\textunderscore LIR& \lsun& Total $8-1000$ $\mu$m \lir\
                                        \citep{draine_2007, mullaney_2011}\\
    (d)SF\textunderscore LIR& \lsun& $L_{\rm IR,\,SFR}$ from the
                                     star-forming component ($L_{\rm
                                      IR,\,SFR}=L_{\rm IR}-L_{\rm IR,\,AGN}$)\\
    (d)AGN\textunderscore LIR& \lsun& $L_{\rm IR,\,AGN}$ from the AGN
                                      component ($L_{\rm
                                      IR,\,AGN}=L_{\rm IR}\times f_{\rm AGN})$\\
    (d)$f_{\rm AGN}$& \dots& Fraction of \lir\ due to the AGN emission\\
    (d)$M_{\rm dust}$& \msun& Dust mass \citep{draine_2007}\\
    (d)$U$& \dots& Mean intensity of the interstellar radiation field \citep{draine_2007}\\
    DistanceMS & \dots& Distance from the main sequence as
                        parameterized in \cite{sargent_2014}\\
    (d)Size& arcsec& Source angular size from ALMA\\
    OneSigma\textunderscore Size& arcsec& $1\sigma$ upper limit on the source angular size from ALMA\\ 
    Probability\textunderscore Unresolved& & Probability of being
                                             unresolved (\citealt{puglisi_2019})\\ 
    \\
    Flux\textunderscore Line($X$)& \jykms& Velocity integrated flux
    of line $X$\\
    SNR\textunderscore ($X$)& \dots& Signal to noise ratio of the
                                         flux of line $X$\\
    OneSigma\textunderscore ($X$)& \jykms& $1\sigma$ upper limit on
                                           the flux of line $X$\\
    Width\textunderscore ($X$)& \kms& Velocity width of line $X$\\
    (d)ApertureCorr\textunderscore ($X$)& \dots& Aperture correction
                                                 for line $X$\\
    Prob\textunderscore Line($X$)& \dots& Probability of spurious
                                          detection of line $X$\\
    Flag\textunderscore ($X$)& \dots& Quality and usage flag for line $X$\\
    FreqContinuumBand($B$)\textunderscore ($X$)& GHz& Frequency in
                                                      Band $B$ under
                                                      line $X$ for the estimate of the continuum emission\\
    (d)ContinuumBand($B$)\textunderscore ($X$)& mJy& Continuum
                                                     emission in Band
                                                     $B$ under line $X$\\
    \midrule
    \bottomrule
  \end{tabular}
  \tablefoot{\smallskip This full table is available in .fits format in
  the online version of this article.
  
  Lines $X$: \cofive, \cotwo, \coseven, \citwo, \cione, and \cofour.\\
  Bands $B$: Band 6, Band 3, and Band 7.\\
  The uncertainties have the same name of the quantity that they refer to,
  preceded by $d$ (e.g., Total\textunderscore LIR $\pm$ dTotal\textunderscore LIR).}
  \label{tab:datatable}
\end{table*}
\end{document}